\documentclass[11pt]{JHEP3}

\usepackage{amsmath,amsfonts,amssymb}
\usepackage{epsf}
\newcommand{\be}{\begin{equation}}
\newcommand{\ee}{\end{equation}}
\newcommand{\beq}{\begin{eqnarray}}
\newcommand{\eeq}{\end{eqnarray}}
\newcommand{\bea}[2]{\be\label{#2}\begin{array}{#1}}
\newcommand{\eea}{\end{array}\ee}

\def\zb{\bar{z}}

\def\sign{{\rm sign}}

\def\({\left(}
\def\){\right)}
\def\[{\left[}
\def\]{\right]}
\def\p{\partial}

\def\11{1\!\! 1}

\def\hf{\frac{1}{2}}

\def\e{\epsilon}

\def\vrh{\varrho}
\def\vph{\varphi}

   \def\CL {{\cal L}}

\def\bz{\bar z}

\def\tQ{{\tilde Q}}

\def\gst{g_{\rm s}}
\def\Qch{Q_{\chi}}
\def\Qph{Q_{\vph}}
\def\Qs{Q_{\sigma}}

\def\Qf{\tQ_5}

\newcommand{\bR}{\mathbb{R}}
\newcommand{\bZ}{\mathbb{Z}}
\newcommand{\Nint}{\mathbb{N}}

\title{Membrane and fivebrane instantons from quaternionic geometry}
\author{ Sergei Alexandrov$^{1,2}$, Frank Saueressig$^{2}$, and Stefan Vandoren$^{2}$ \\
$^{1}$Laboratoire de Physique Th\'eorique \& Astroparticules \\
Universit\'e Montpellier II, 34095 Montpellier Cedex 05, France \\
\\
$^{2}$Institute for Theoretical Physics \& Spinoza Institute \\
Utrecht University, Postbus 80.195, 3508 TD Utrecht, The Netherlands\\
\\
E-mail: \email{S.Alexandrov, F.S.Saueressig, S.Vandoren@phys.uu.nl}}
\abstract{We determine the one-instanton corrections to the
universal hypermultiplet moduli space coming both from Euclidean
membranes and NS-fivebranes wrapping the cycles of a (rigid)
Calabi-Yau threefold. These corrections are completely encoded by a
single function characterizing a generic four-dimensional
quaternion-K\"ahler metric without isometries.
We give explicit solutions for this function
describing all one-instanton corrections, including the fluctuations
around the instanton to all orders in the string coupling constant.
In the semi-classical limit these results are in perfect agreement with
previous supergravity calculations.}
\preprint{LPTA-06/16 \\ ITP-UU-06/29 \\ SPIN-06/25 \\ hep-th/0606259}


\begin{document}

\section{Introduction}

Recently there has been considerable progress in understanding the quantum
corrections to the hypermultiplet moduli spaces arising from compactifying
type II strings on Calabi-Yau threefolds (CY$_3$). These
moduli spaces appear as sigma models for hypermultiplets
in the $N=2$, $D=4$ low-energy effective supergravity action. Since for type
II strings the dilaton lives in a hypermultiplet, this sector is subject
to both perturbative and non-perturbative stringy corrections.

In general the real dimension of the hypermultiplet
moduli space is given by $4 n_H$ where the number of hypermultiplets
is given by $n_H = h_{1,2}+1$ for type IIA and $n_H = h_{1,1}+1$
for type IIB, where $h_{i,j}$ are the Hodge numbers of the CY$_3$.
Local supersymmetry implies that the hypermultiplet moduli space
has to be quaternion-K\"ahler \cite{Bagger:1983tt},
i.e., for $n_H > 1$ the space has to be Einstein with holonomy
contained in Sp(1) $\times$ Sp($n_H$). The  case $n_H = 1$ is special.
Here the definition of quaternion-K\"ahler geometry implies that
the hypermultiplet moduli space has to be Einstein with
self-dual Weyl curvature. This situation appears in type IIA compactifications 
on rigid CY$_3$, for which the Hodge number $h_{1,2}=0$, and 
corresponds to the universal hypermultiplet.

At tree-level, the metric on the hypermultiplet moduli space can be
determined either through the c-map
\cite{Cecotti:1988qn,Ferrara:1989ik,Rocek:2005ij} or by explicit dimensional 
reduction from ten dimensions
\cite{BCF,BGHL}. 

In perturbation theory, the hypermultiplet moduli space admits a number
of commuting isometries which simplifies the description of the underlying 
quaternionic geometry. In that case, one can use the off-shell formulation
for tensor multiplets which describe $4n_H$-dimensional quaternion-K\"ahler
manifolds with $n_H+1$ commuting isometries \cite{deWit:2006gn}. This framework
allowed to determine the perturbative one-loop corrections for
generic Calabi-Yau compactifications
\cite{Robles-Llana:2006ez}, building on earlier work
\cite{Antoniadis:1997eg}, and generalizing the case of the universal
hypermultiplet \cite{AntoniadisUH,Anguelova:2004sj}. It was further argued in
\cite{Robles-Llana:2006ez} that a non-renormalization theorem
protects the hypermultiplet moduli space metric from higher loop
corrections.

Non-perturbatively, there can be spacetime instanton corrections coming from
wrapped Euclidean branes over the relevant cycles of the CY$_3$ \cite{BBS}.
These will generically break (some of) the isometries present in perturbation
theory. Beyond that, not much is known about their contribution to the 
hypermultiplet moduli
space in the generic case with $h_{1,2}\neq 0$. This is due to our
poor understanding of non-perturbative string theory, as well as our limited
knowledge of the quaternion-K\"ahler geometry that underlies the
hypermultiplet moduli space, see e.g.
\cite{Aspinwall:2000fd} for a general discussion.
For the universal hypermultiplet, however, there are some partial results
either for membrane instantons \cite{DSTV}, or for NS-fivebrane instantons
\cite{Davidse:2004gg}. These results grew out of earlier work
\cite{Theis:2002er,DdTV,Anguelova:2004sj}, see also
\cite{BB2,GS1,Ketov:2001gq,Ketov:2002vr}. Recently, these instanton
corrections also played a prominent role in the construction of 
meta-stable de Sitter vacua \cite{DSTV,STV}.

The aim of this paper is to determine the non-perturbative
corrections to the universal hypermultiplet moduli space metric,
including both membrane and fivebrane instantons. The key property
for studying such corrections is that the corresponding metric must
be quaternion-K\"ahler. Such metrics have been
studied extensively in the context of Euclidean relativity and it
was found in \cite{Prz3} that the general metric can be encoded in a
single function satisfying a non-linear partial differential
equation. We then find solutions to this equation which describe
generic instanton corrections. As a non-trivial test, we reproduce
the formulae for the instanton actions calculated in the
supergravity approximation in \cite{Theis:2002er,DdTV}. Moreover,
our results are in perfect agreement with the explicit fivebrane
instanton calculations done in the supergravity approximation
\cite{Davidse:2004gg}. In addition, for both membrane and fivebrane cases,
we analyze the contributions
from the fluctuations around the one-instanton to all orders in the
string coupling constant. In particular,
this includes the one-loop determinant around the instanton. It is remarkable
that the constraints from quaternionic geometry, combined with
sensible boundary conditions, are restrictive enough to determine
the form of the non-perturbative corrections that arise in string theory.

The rest of the paper is organized as follows. In Section 2 we
review the classical universal hypermultiplet moduli space and its
perturbative corrections. In Section 3 we present the metric for a
generic four-dimensional quaternion-K\"ahler manifold in terms of a
single function $h$ and determine this function for the universal 
hypermultiplet and its perturbative corrections. 
Then, in Section 4, we discuss the
one-instanton corrections to the function $h$, together with the
perturbative fluctuations around it. In particular we obtain explicit 
expressions for instanton corrections due to membranes and fivebranes and
show their agreement with previously
known results obtained in the semi-classical supergravity limit. We end
with a brief discussion of our results in Section 5. The
technical details of our calculations can be found in the
Appendices.

\section{The perturbative universal hypermultiplet}

We start by reviewing the results for the perturbatively corrected
universal hypermultiplet. This theory arises from compactifying type
IIA strings on a rigid CY$_3$. The hypermultiplet sector of the
4-dimensional low energy effective Lagrangian can be obtained from
dimensional reduction. At tree-level the relevant bosonic terms read
\be
\begin{split}
e^{-1} \CL_{\rm T} = & \, - R - {\tfrac12} \,\partial_\mu \phi
\partial^\mu \phi + {\tfrac12}\, {\rm e}^{2 \phi} H_\mu H^{\mu} \\
& \,  -
{\tfrac12}{\rm e}^{-\phi}(\partial_\mu \chi \partial^\mu \chi +
\partial_\mu \varphi \partial^\mu \varphi)
- {\tfrac12} H^{\mu} (\chi \partial_\mu \varphi
- \varphi \partial_\mu \chi) \, .
\end{split}
\ee
Here $\phi$ is the four-dimensional dilaton\footnote{Our conventions
are such that the string coupling constant is related to the
asymptotic value of the dilaton via $\gst={\rm
e}^{-\phi_{\infty}/2}$.}, $H^\mu = \tfrac1{6} \varepsilon^{\mu \nu
\rho \sigma} H_{\nu \rho \sigma}$ is the NS two-form field strength,
and $\varphi$ and $\chi$ can be combined into a complex scalar $C$
that descends from the holomorphic components of the RR 3-form with
(complex) indices along the holomorphic 3-form of the CY$_3$. The NS
two-form can then be dualized to an axion $D$ by introducing a
Lagrange multiplier
\be
e^{-1} \CL_{\rm LM} = - H^\mu \p_\mu D \, ,
\ee
and eliminating $H^\mu$ through its equation of motion. From the supergravity
point of view it is often convenient to redefine this axion using the field
\be
\sigma = D - \tfrac12 \chi \vph \,.
\label{Dsigma}
\ee
In terms of $\chi, \vph, \sigma$ and $r = {\rm e}^{\phi}$ the classical
universal hypermultiplet Lagrangian becomes
\be \label{HM}
e^{-1} \CL_{\rm UHM}^{\rm cl} = - R - \frac{1}{2r^{2}}\,
\left[
(\p_{\mu} r)^2 + r \( (\p_\mu \chi)^2  + ( \p_\mu \vph
)^2 \) + \( \p_\mu \sigma+\chi \p_\mu \vph\)^2 \, \right] .
\ee
The sigma model target space of the classical universal hypermultiplet
is $SU(2,1)/U(2)$ \cite{Cecotti:1988qn,Ferrara:1989ik}, with isometry group
$SU(2,1)$.

The perturbative corrections to the universal hypermultiplet have recently
been obtained in
\cite{AntoniadisUH}. There it was found that the Lagrangian \eqref{HM} receives a non-trivial
one-loop correction while higher loop contributions can be
absorbed by a coordinate transformation. The perturbatively corrected
hypermultiplet metric then reads
\be
{\rm d}s_{{\rm UHM}}^2=\frac{1}{r^2}\[ \frac{r+2c}{r+c}\, {\rm d}r^2+
(r+2c)\({\rm d}\chi^2+{\rm d}\vph^2\)+
\frac{r+c}{r+2c}\( {\rm d}\sigma+\chi {\rm d}\vph\)^2\]\ .
\label{HMq}
\ee
The constant $c$ was obtained as
\be
c=-\frac{4\zeta(2)\chi(X)}{(2\pi)^3}=-\frac{1}{6\pi}(h_{1,1}-h_{1,2})\ ,
\ee
with $\chi(X)$ denoting the Euler number of the CY$_3$, and one should
set $h_{1,2}=0$ in our case.

At the perturbative level the hypermultiplet metric \eqref{HMq} retains four
unbroken isometries. First of all there is a 3-dimensional Heisenberg group
of isometries acting as shifts in $\sigma$, $\chi$ and $\vph$
\be
r\to r,\qquad \chi\to\chi+\gamma,
\qquad \vph\to\vph+\beta, \qquad \sigma\to \sigma
-\alpha-\gamma\vph \, ,
\label{HMisom}
\ee
where $\alpha, \beta$ and $\gamma$ are real parameters. Additionally one has
a rotational symmetry in the $\chi$-$\vph$ plane parameterized by the real
angle $\delta$
\be \label{HMis}
 r\to r, \qquad
 D \to D, \qquad \chi\to\chi\cos\delta-\vph \sin\delta, \qquad
\vph\to\chi\sin\delta+\vph \cos\delta\  \, .
\ee
In this frame it is obvious that the isometry \eqref{HMis}
corresponds to a phase transformation of the complex coordinate 
$C=\frac{1}{2}(\chi+i\varphi)$
which, microscopically, is related to a rescaling of the Calabi-Yau
holomorphic three-form by a phase. Notice that when working in terms
of $\sigma$ the isometry \eqref{HMis} acts non-trivial on $\sigma$:
\be
\begin{split}
 \sigma\to \sigma +\chi\vph \sin^2\delta+
\tfrac{1}{4}\,(\vph^2-\chi^2) \sin(2\delta).
\end{split}
\ee
Generically, one expects that (some of) these isometries will be broken by
instanton corrections. For instance, fivebrane instantons will break
the shift symmetry in $\sigma$ to a discrete subgroup, whereas
membrane instantons will break the shift symmetry in $\varphi$ or $\chi$.
We will discuss the fate of these isometries in more detail in
Section 4.

\section{Four-dimensional quaternion-K\"ahler geometry}

After having reviewed the perturbatively corrected universal
hypermultiplet metric, we now proceed and
introduce the Przanowski framework describing a general
four-dimensional quaternion-K\"ahler metric.
Subsequently we will recast the perturbative universal
hypermultiplet in this framework and derive the linear partial
differential equation governing small perturbations around the
perturbative metric in subsections 3.2 and 3.3, respectively.

\subsection{The master equation}

Four-dimensional quaternion-K\"ahler spaces coincide with Riemannian
Einstein manifolds with a self-dual Weyl tensor. Such manifolds with
a non-vanishing cosmological constant $l$ were characterized
in \cite{Prz1,Prz2,Prz3} in terms of solutions of a single differential
equation. More precisely, it has been shown that there always exists
a system of local complex coordinates, $z^1$ and $z^2$, such that
the metric takes the form \be {\rm d}s^2=g_{\alpha\bar\beta}\,({\rm
d}z^{\alpha} \otimes  {\rm d}z^{\bar\beta}+ {\rm d}z^{\bar\beta}
\otimes {\rm d}z^{\alpha})\ , \ee where $\alpha,\beta=1,2$,
$z^{\bar\beta}=\bar z^\beta$ and \be
g_{\alpha\bar\beta}=-\frac{3}{l}\(\p_{\alpha}\p_{\bar\beta}h+
2\delta_{\alpha}^2\delta_{\bar \beta}^{\bar 2}\, {\rm e}^h \)\ .
\label{metPrz} \ee This form for the metric components was also
derived more recently in \cite{deWit:2001dj}, where it was
generalized to quaternion-K\"ahler spaces of higher dimensions.
Notice that the metric is completely characterized in terms of a
single real function $h(z^{\alpha},z^{\bar\alpha})$. The constraints
from quaternionic geometry imply \cite{Prz1} the following
non-linear partial differential equation \be \p_{1}\p_{\bar
1}h\cdot\p_{2}\p_{\bar 2}h-\p_{1}\p_{\bar 2}h\cdot\p_{\bar 1}\p_{
2}h +\(2\p_{1}\p_{\bar 1}h-\p_{1}h\cdot\p_{\bar 1}h\){\rm e}^h=0\ .
\label{master} \ee We call \eqref{master} the master equation. It
describes in a compact way all four-dimensional quaternion-K\"ahler
geometries.

Furthermore, in \cite{Prz3} this general result was specialized to
the case of quaternion-K\"ahler metrics with one Killing vector
field. It turns out that there are two distinct situations depending
on the direction of the isometry. According to \cite{Prz3} there can
be shifts along either \be {\rm (A)}\quad z^2-z^{\bar 2},\qquad {\rm
or}\qquad {\rm (B)}\quad z^1-z^{\bar 1}. \label{isom} \ee In both
cases the metric is completely determined by the solutions of
\eqref{master}, which do not depend on the combination \eqref{isom}.

For the case (B), it was shown in \cite{Prz3} that the master
equation \eqref{master} reduces to the three-dimensional Toda
equation \be \p_z\p_{\bar z}F+\p_r^2 {\rm e}^F=0, \label{Toda} \ee
by means of the Lie-B\"acklund transformation \be z=z^2,\quad \bar
z=\bar z^2,\quad r=-\(\p_{z^1+z^{\bar 1}}h\)^{-1},\qquad F=h-\log
\(\p_{z^1+z^{\bar 1}}h\)^2 \, . \ee For a related discussion on such
geometries, see also \cite{Tod}. So all self-dual Einstein metrics
possessing a Killing vector of type (B) can be characterized by
solutions of the three-dimensional Toda equation. For the case (A)
no such transformation is known explicitly, although the results of \cite{Tod}
imply this is still possible.\footnote{We thank P. Tod for a discussion
on this issue.} 

It is somewhat surprising that the two classes in \eqref{isom} have some 
physical meaning. For fivebrane instantons only, one can work in 
a framework in which the isometry of type (A) is manifest. In the 
case of membrane instantons only, i.e., in the
absence of fivebrane charge, the isometry
of type (B) is preserved. This follows from the results of the next 
subsection. The general situation is however 
when both types of instantons are present and hence no isometries will be
preserved. This is the situation we describe in this paper, and we will 
therefore stick to the master equation
\eqref{master} and its solutions.

\subsection{The universal hypermultiplet in the Przanowski framework}

When using the Przanowski metrics above as target spaces in a
low-energy supergravity action, supersymmetry requires to fix the
value of the Ricci-scalar constructed from these metrics in terms of
the gravitational coupling constant $\kappa$. In our conventions
where we have chosen $\kappa^{-2} = 2$, the proper value is $R =
-6$, implying that the value of the cosmological constant is fixed
to $l=-\frac{3}{2}$.

To recast the perturbatively corrected universal hypermultiplet
metric into Przanowski form, one should find a coordinate
transformation which maps \eqref{HMq} to the metric \eqref{metPrz}
with the function $h$ satisfying \eqref{master}. This is achieved by
setting
\be z^1=\hf(u+i\sigma),\qquad z^2=\hf(\chi+i\vph), \label{change}
\ee where
\be
u=r-\hf\chi^2+c\log(r+c) \label{uy} \, .
\ee
The corresponding function $h_0$ is given by
\be
h_0=\log(r+c)-2\log r \, ,
\label{hHM}
\ee
and we have added the subscript $0$ in order to indicate the
perturbative solution. In this formula, $r$ has to be understood as
a function of $z^1$ and $z^2$. The inverse transformation of
\eqref{uy} can only be done explicitly for $c=0$. In that case, the
classical solution is given by $h_0=-\log r$ with
\begin{equation}
c=0\,: \quad r = z^1 +\bar{z}^1 +\frac{1}{2}(z^2+\bar {z}^2)^2\ .
\end{equation}
It is easy to verify that this solves \eqref{master}.
To verify that the function \eqref{hHM} is a solution of
\eqref{master} it is
useful to first rewrite the master equation in terms of the real
fields $u,\sigma,\chi$ and $\vph$,
\begin{eqnarray}\label{Przreal}
&&(\p_{\chi}^2+\p_{\vph}^2)h\,(\p_u^2+\p_{\sigma}^2)h
-(\p_{\chi}\p_u h+\p_{\vph}\p_{\sigma} h)^2-
(\p_{\chi}\p_{\sigma} h-\p_{\vph}\p_u h)^2  \nonumber\\
&& \, +\Big(2(\p_u^2+\p_{\sigma}^2)h-(\p_u
h)^2-(\p_{\sigma}h)^2\Big){\rm e}^h=0.
\end{eqnarray}
Changing basis again to the variables $r,\sigma,\chi$ and $\vph$
using the identities \be\label{eq:IdA}
\begin{split}
\p_u = \frac{r+c}{r+2c} \p_r \, ,
\quad
\left.\p_{\chi}\right|_{u}= \left.\p_{\chi}\right|_{r}+\chi \frac{r+c}{r+2c} \p_r \, , \quad
 \p_{u} h_0=-\frac{1}{r} \, , \quad
\p_{\chi} h_0= - \frac{\chi}{r} \, ,
\end{split}
\ee it is then straightforward to check that \eqref{hHM} solves the
master equation.

Moreover, note that $h_0$ is a function of $u$
and $\chi$, and thus of $z^1+\zb^1$ and $z^2+\zb^2$ only.
Hence there are shift symmetries associated with both $z^1-\zb^1$
and $z^2-\zb^2$ such that the perturbative hypermultiplet metric
fits simultaneously into both cases \eqref{isom}. In particular,
this means that it can be described by a solution of the Toda
equation \eqref{Toda}. This fact was extensively used in \cite{DSTV}
when studying membrane instanton corrections to the universal
hypermultiplet. Performing the Lie-B\"acklund transformation for the
solution \eqref{hHM}, we obtain ${\rm e}^F=r+c$ which 
reproduces the result found in \cite{DSTV}.

\subsection{Deformations around the perturbative solution}

After having established the Przanowski description of the perturbatively 
corrected
universal hypermultiplet we now proceed and discuss the inclusion of
non-perturbative corrections. In this course we will work in the region
of the moduli space where the string coupling is small, so that these
corrections are
exponentially suppressed. In terms of the real variables
this means that $r$ and $u$ are much larger than any other variable.

The solution describing the non-perturbative corrections
to the universal hypermultiplet can be written as
\be
h=h_0(u,\chi)+\Lambda(u,\sigma,\chi,\vph) + \ldots \, .
\label{hcor}
\ee
Here $h_0(u,\chi)$ is the perturbative solution \eqref{hHM} and $\Lambda$
is an exponentially small correction. We assume that it encodes
the one-instanton effects, whereas multi-instanton corrections
are hidden in the dots.
They will not be considered in the following and therefore
we do not display them explicitly in our ansatz. We will therefore
work in the linearized approximation and assume that the solution can be
extended to a solution of the full non-linear equation by setting up 
an iteration scheme. For the case of the Toda equation, this was demonstrated
in \cite{DSTV}.

Utilizing \eqref{Przreal} we can derive a partial differential equation
for $\Lambda$ which implements the constraints from quaternion-K\"ahler
geometry. Substituting the ansatz \eqref{hcor} into \eqref{Przreal} and expanding the result
to the first order in $\Lambda$ one finds
\beq
&\(\p_{\chi}^2 h_0+2{\rm e}^{h_0}\)(\p_u^2+\p_{\sigma}^2)\Lambda
+\p_{u}^2 h_0 (\p_{\chi}^2+\p_{\vph}^2)\Lambda
-2\p_{\chi}\p_u h_0 (\p_{\chi}\p_u+\p_{\vph}\p_{\sigma})\Lambda
&\nonumber \\
& -2{\rm e}^{h_0}\p_u h_0 \p_u\Lambda
+{\rm e}^{h_0}\( 2\p_u^2 h_0-(\p_u h_0)^2\)\Lambda=0. &
\label{Leqfull}
\eeq
For our purpose it is, however, more convenient to work directly in terms of
the hypermultiplet variables. Therefore, we trade the variable $u$ for $r$ by means of \eqref{uy}.
Then using the formulas \eqref{eq:IdA}, one can rewrite \eqref{Leqfull} as
\be
\[\(r+\chi^2+3c+\frac{c^2}{r+c}\)\p_{\sigma}^2
+(r+c)\p_r^2+\p_{\chi}^2+\p_{\vph}^2 -2\chi\p_{\vph}\p_{\sigma}
+\(3+\frac{2c}{r}\)\p_r+\frac{1}{r}\]\Lambda=0.
\label{Leq}
\ee
This equation is the master equation for the instanton
corrections.

\subsection{Solution generating technique}
\label{ss:3.4}

Before we discuss the solutions of \eqref{Leq} in the next section, let 
us first point out a general solution generating technique \cite{Prz4}.
The master equation \eqref{master} has an invariance group of transformations,
as one can easily verify,
\begin{equation}
z^1\rightarrow {\tilde z}^{1} = f(z^1,z^2)\ ,\qquad z^2 \rightarrow
{\tilde z}^{2} = g(z^2)\ ,
\end{equation}
where $f$ and $g$ are two arbitrary holomorphic functions. These
transformations lead to new solutions of \eqref{master} of the form
\begin{equation}
h(z^1,z^2) \rightarrow {\tilde h}(z^1,z^2) =
h\Big(f(z^1,z^2),g(z^2)\Big) - \log (g'(z^2)
{\bar g}'(\bar {z}^2))\ ,
\end{equation}
where the prime stands for the derivative.

In general these new solutions lead to new metrics with different asymptotics.
In our case, we want the new metric to have the same asymptotic
behavior, namely the one given by \eqref{hHM}. This puts certain constraints
on $f$ and $g$, and a class of functions which leave the boundary
conditions invariant are the ones that generate the isometries. For instance,
choosing
\begin{equation}
f(z^1,z^2)=z^1-\gamma z^2-\frac{1}{4}\gamma^2-\frac{1}{2}i\alpha \ ,
\qquad
g(z^2)=z^2+\frac{1}{2}(\gamma+i\beta)\ ,
\end{equation}
generates the Heisenberg group of isometries \eqref{HMisom}, and similarly
for the rotations \eqref{HMis}. Clearly, the
solution $h_0$ as in \eqref{hHM} is invariant, and this guarantees
that the asymptotic form remains the same. On the other hand, the instanton
deformations $\Lambda$ from \eqref{hcor} will not be invariant since some
of the isometries from the Heisenberg group are broken non-perturbatively.
However, at the level of the hypermultiplet metric the new
solutions are related by a coordinate transformation. Thus, the
solutions generated in this way do not produce physically
inequivalent spaces.

This feature can be overcome when one works at the linearized level.
It is clear that the perturbative isometries also generate new
solutions of the linearized eq.\ \eqref{Leq}. But in contrast to
the non-linear case, this technique allows to generate new target
spaces by taking linear combinations of the transformed instanton
corrections.
For example, having a solution $\Lambda_0(r, \chi, \vph, \sigma)$ of
\eqref{Leq}, the transformation $\vph \rightarrow \vph + \beta$
generates the family of solutions
\be
\Lambda_\beta(r, \chi, \vph, \sigma) = \Lambda_0(r, \chi, \vph + \beta, \sigma).
\ee
All of them are physically equivalent. But the superposition of instanton
solutions of the form
\be
\int {\rm d}\beta\, C(\beta)\Lambda_\beta(r, \chi, \vph, \sigma)
\ee
with any function $C(\beta)$ having support on more than one point,
generically leads to physically distinct target spaces.

These facts can be used to simplify the derivation of the general
instanton solution, since they allow us to focus on one particular
member of a family of solutions related by the isometry
transformations. Once we find its exact form, all other solutions
will follow by applying the perturbative isometry transformations
and considering their linear combinations.

\section{Instanton corrections to the universal hypermultiplet}

\subsection{Supergravity results about instanton actions}

Before we start the calculation of instanton corrections to the
universal hypermultiplet, let us briefly review the results
obtained in the supergravity framework about the instanton
actions \cite{Theis:2002er,DdTV,Davidse:2004gg}.

There are two classes of instanton solutions, describing the wrapping
of a NS-fivebrane or a membrane in type IIA string compactifications on a
rigid CY$_3$. They produce corrections to the metric
proportional to ${\rm e}^{-1/\gst^2}$ or ${\rm e}^{-1/\gst}$ respectively
\cite{BBS}, and we repeat (see footnote 1) that the relation with the
dilaton is $g_s={\rm e}^{-\phi_{\infty}/2}$. These instantons
can also be described as finite action solutions to the four-dimensional
supergravity equations of motion for the universal hypermultiplet
\cite{Theis:2002er,DdTV}, see also \cite{GS1} for an earlier reference.

It was found that, for the NS-fivebrane instantons, the supergravity
instanton action is given by\footnote{Actually, in
\cite{Theis:2002er,DdTV} the instanton action was written in terms of
$\Delta\chi=\chi-\chi_0$ and $\hat\sigma=\sigma +\chi_0\vph$, where
$\chi_0$ is an arbitrary constant that was interpreted as a RR flux.
It is clear that this result for the instanton action can be
obtained  from \eqref{fiveinst} by applying the shift isometry in
$\chi$ \eqref{HMisom}, with parameter $\gamma=-\chi_0$. Finding a
solution of the master equation consistent with \eqref{fiveinst} is
therefore sufficient. Similarly for the membrane instantons. We come back to
this issue in
subsection 4.3 \label{footn}} \be S^{(5)}_{{\rm
inst}}=|Q_5|\(\frac{1}{\gst^2}+\hf\chi^2\) +iQ_5\sigma. \label{fiveinst} \ee
Here, $\chi$ is treated as a coordinate on the moduli space,
similarly to the dilaton. It is natural to interpret $Q_5$ as the
instanton number for the Euclidean NS-fivebrane, wrapped over the
entire Calabi-Yau space. The $\theta$-angle like term ${\rm e}^{iQ_5\sigma}$ 
breaks the isometry of the classical metric along $\sigma$ to a discrete
subgroup $\bZ$. Notice further that the shift symmetry in $\varphi$
is unbroken so, in the presence of fivebrane instantons only, this
remains an isometry. Finally, the shift symmetry in $\chi$ is
explicitly broken.

The membrane instantons, as found in \cite{Theis:2002er,DdTV,Davidse:2004gg}, 
can be parameterized by two charges and the instanton
action is \be S^{(2)}_{{\rm inst}}=\(|Q_2|+\hf|\chi
Q_5|\)\sqrt{\frac{4}{\gst^2}+\chi^2}+iQ_2\varphi+iQ_5\sigma
. \label{meminst} \ee For
$\chi=0$ one obtains a more standard instanton action inversely
proportional to the string coupling constant, $S=2|Q_2|/g_s$, in
which $Q_2$ plays the role of the membrane charge. 
Notice that they also contain the fivebrane
charge $Q_5$. The microscopic
string theory interpretation of the additional terms proportional to
$\chi$ remains unclear. 
In the next subsections, we will show however that
there exist solutions to the master equation \eqref{Leq} that
reproduce these results exactly. 

The fate of the isometries for membrane instantons differs from the 
fivebrane case, in the sense 
that the isometry along $\vph$ is also broken to a discrete
subgroup. When $Q_5$ is switched off, the shift along
$\sigma$ remains to be an isometry.

\subsection{General constraints on instanton corrections}
\label{masterinst}

Our main goal is to determine the instanton corrections to the
perturbative hypermultiplet metric \eqref{HMq} using the same
strategy as in \cite{DSTV}. Namely, we find exponentially small
corrections to the solution \eqref{hHM} of the master equation
\eqref{master}. These corrections in turn generate corrections to
the metric, and by the results of \cite{Prz1,Prz2} the full metric
will automatically satisfy the constraints of the quaternionic
geometry. The main difference with \cite{DSTV}, based on solutions of the Toda
equation, is that we do not suppose the existence of an isometry since
the combination of both membrane and fivebrane instantons does not
preserve any continuous isometry.

Our goal is to solve the master equation \eqref{Leq} and
consequently determine the instanton corrected hypermultiplet moduli
space metric. Of course, this master equation possesses a large
number of solutions. We are not interested in all of them but only
in those which are physically appropriate. Therefore, we impose a
set of requirements on the solution to be satisfied. Our conditions
on admissible instanton corrections are the following:
\begin{enumerate}
\item For small string coupling $\gst$ the instanton
contributions should be exponentially suppressed.
\item In each instanton sector, there should be a perturbative
series in $\gst$ that describe the fluctuations around the
instanton. These we represent as an (exponentiated) Laurent series,
bounded from below and including the term $\log\gst$.

\item The shift symmetry in the NS scalar $\sigma$ (or $D$) should be broken
by a theta-angle-like term only. This requirement is justified by
the fact that the NS scalar comes from dualizing the NS two-form in
four dimensions. The breaking of the isometry is triggered by the
boundary term in the dualization process. This boundary term
precisely generates the theta-angle that breaks the shift symmetry
to a discrete subgroup.

\item For instanton solutions containing several charges, for instance 
the membrane and
fivebrane instanton charges, the limit where one of the charges
vanishes should still give rise to a regular solution which is
exponentially suppressed.

\item The theta-angle terms should be independent of $\gst$.
By this we mean that the purely imaginary terms in the exponent are
independent of $\gst$

\end{enumerate}

Some of these requirements reflect a more stringent condition, namely
that the full non-perturbative solution leads to a {\em regular} metric on
the hypermultiplet moduli space. The perturbatively corrected 
metric corresponding to \eqref{hHM} develops a singularity at $r=-c$, and 
the instantons are supposed to resolve this singularity. This resolution
can however not be understood in the one-instanton approximation we are working
in. One would need to determine and sum up the entire instanton series to
see how the singularity gets resolved. For an example of how this can
work, we refer to \cite{OV}.

An instanton correction which satisfies the first two conditions can
be written in the following general form
\be
\Lambda= A_0
r^{\alpha}\exp\(-\sum_{k=1/2}^p f_{k} r^{k}
+\sum_{k\in\Nint/2}^{\infty}\frac{A_{k}}{r^{k}} \). \label{genans}
\ee
The terms containing the $f_k$ are proportional to powers
of the inverse string coupling constant $\gst^{-1}={\sqrt r}$. They
reflect the non-perturbative nature of the solution. The terms
containing the $A_k$ describe the fluctuations around the instanton.
Usually, these are written in terms of a power series in $\gst$ in
front of the exponent, but they can be exponentiated as in
\eqref{genans} up to the term $r^\alpha$ that would lead to a
logarithm in the exponent. Here $\alpha$ is some constant, $p\in
\Nint/2$, and the sums over $k$ run over (positive) integers and
half-integers. All the coefficients, including $A_0$, are complex
functions of $\chi$, $\varphi$ and $\sigma$. The solution we
construct is therefore complex, but we can add the complex conjugate
to obtain a real solution. These two sectors will describe
instantons and anti-instantons respectively. To leading order, which
is the approximation we are working in, there is no mixing between
these two sectors.

However, combining the conditions (3) and (5), one immediately concludes
that all $f_k$ and $A_k$ except $A_0$ must be real and
$\sigma$-independent. We further study this generic ansatz in
appendix \ref{D} where we demonstrate that
\begin{equation}
\forall k>1 \quad :\quad f_k=0\ .
\end{equation}
We also prove there, in the linearized approximation we are
working in, that there are no solutions with both $f_1$ and
$f_{1/2}$ non-vanishing. Such terms might be generated however at
subleading order, and would correspond to a combined system where both
membrane and fivebrane instantons are present.

Thus, in accordance with the supergravity result, there are two
classes of instanton corrections satisfying our conditions, which
scale as ${\rm e}^{-f_1/\gst^2}$ and ${\rm e}^{-f_{1/2}/\gst}$,
respectively. These two classes are clearly related to NS-fivebrane
and membrane instantons respectively \cite{BBS}.
As is known in string theory and will be confirmed by solving
the master equation, in the first case the perturbative expansion around the
instanton goes in even powers of $\gst$, whereas in the second case
all integer powers contribute.

In the following we discuss the most general solutions, up to the
action of the isometry group (see the end of section 3.3) of
\eqref{Leq} satisfying all the above requirements which fit in these
two classes.

\subsection{Fivebrane instantons}

In this subsection we study instanton corrections arising due to a
Euclidean NS-fivebrane wrapping the entire Calabi-Yau. From the
supergravity analysis it is known that the corresponding instanton
action scales like $\gst^{-2}$ and such corrections fit to our
solutions with non-vanishing $f_1$. Since the perturbative expansion
around such an instanton should go in even powers of $\gst$, one
arrives at the following ansatz \be \Lambda^{(5)}=A_0
r^{\alpha}\exp\(-f_1 r+\sum_{k\in \Nint}^{\infty}\frac{A_{k}}{
r^{k}}\). \label{fiveans} \ee

In appendix \ref{B} we solve the master equation \eqref{Leq} using
the ansatz \eqref{fiveans} and subject to the constraints listed
above. We find the following {\it exact} solution\footnote{In fact,
 we can find more general 
solutions where the function $Z(r)$ is 
replaced by $Z(r,\rho)$ with $\rho=\chi^2+\varphi^2$. 
All these solutions have the same
semi-classical behavior and differ
in the subleading terms of the $g_s$ expansion
only. In the main text, we have focused on the 
simplest type of solution given by $Z(r)$.
Some more details on the general case can be found in Appendix B.}:
\be
\Lambda^{(5)}= {\rm e}^{\pm
iQ_5\(\sigma+\hf\,\chi\vph\)} \,{\rm
e}^{-\frac{Q_5}{4}\,\(\chi^2+\vph^2\)}\, Z(r)\ .
\label{sol5sym}
\ee
Here, $Q_5$ is strictly positive, $Q_5>0$, and
we denoted
\be
Z(r)=\frac{C\, {\rm e}^{Q_5r}}{ r(r+c)^{cQ_5}}\,
\int_1^{\infty}{\rm e}^{-2Q_5(r+c)t}\frac{{\rm d}t}{t^{1+2cQ_5}}\ ,
\label{ydep}
\ee
where $C$ is some undetermined constant.
Notice that this solution respects the U(1)
isometry \eqref{HMis}. Its leading term in the expansion in powers
of $\gst$ reads
\be
\Lambda^{(5)} \approx C r^{-2-c Q_5}
\,{\rm e}^{\pm iQ_5\(\sigma+\hf\,\chi\vph\)}\,
\exp\[-Q_5\(r+\frac{1}{ 4}\,\(\chi^2+\vph^2\)\)\].
\ee
By expanding $Z(r)$ to higher powers
in $1/r$, one generates the loop expansion around the one-instanton sector
to all orders in the string coupling constant. 

By looking at the form of 
the exponent, we observe that this solution does
not agree with the supergravity result \eqref{fiveinst}. However,
the instanton correction \eqref{sol5sym} is only one member of a
family of solutions related by the perturbative isometry
transformations. In particular, one can use the shift symmetry
\eqref{HMisom} to restore the isometry along $\vph$ which is
manifest in \eqref{fiveinst}. For this let us consider an instanton
correction given by the following integral
\be
\tilde\Lambda^{(5)}\equiv {\textstyle\sqrt\frac{Q_5}{4\pi}}
\int_{-\infty}^{\infty}
\Lambda^{(5)}(r,\sigma,\chi,\vph+\beta)\,{\rm d}\beta =
{\rm e}^{\pm iQ_5\sigma}\,{\rm e}^{-\hf\,Q_5\chi^2}Z(r)\ .
\label{sol51ex}
\ee
The leading term of this
solution now becomes
\be
\tilde\Lambda^{(5)} \approx C r^{-2-c Q_5}
\,{\rm e}^{\pm iQ_5\sigma}\,\exp\[-Q_5(r+\hf\chi^2)\]\ ,
\label{leadsol5}
\ee
and shows precise agreement with the supergravity
result \eqref{fiveinst} for the instanton action.

The isometry transformations allowing to generate new solutions can
be applied as well to the solution \eqref{sol51ex}. For example,
using the $\gamma$-shift symmetry in $\chi$ \eqref{HMisom}, one
generalizes \eqref{sol51ex} to
\be
{\tilde \Lambda}^{(5)}_{\chi_0} =
{\rm e}^{\pm iQ_5(\sigma+\chi_0\vph)} \,
{\rm e}^{-\frac{Q_5}{2}(\Delta\chi)^2} Z(r)\ ,
\label{sol5shift}
\ee
where $\Delta\chi=\chi-\chi_0$ and $\chi_0$ is an arbitrary
constant. This reproduces the instanton solution of \cite{DdTV}
where $\chi_0$ was interpreted as a RR-flux, see also footnote \ref{footn}.

If instead one uses the rotation isometry \eqref{HMis}, one obtains
instanton corrections of the following form
\be
\tilde\Lambda_{\delta}^{(5)} =
{\rm e}^{\pm iQ_5\(\sigma +\chi\vph \sin^2\delta+ \frac{1}{
4}\,(\vph^2-\chi^2) \sin(2\delta)\)} \,{\rm e}^{-\frac{Q_5}{ 2}
\(\chi\cos\delta-\vph \sin\delta\)^2}Z(r).
\label{soldel}
\ee
Notice that for the particular angle $\delta=\pi/2$ the correction
\eqref{soldel} coincides with the one obtained by
integrating \eqref{sol5shift} over $\chi_0$
\begin{equation}
\tilde \Lambda_{\pi/2}^{(5)}={\rm e}^{\pm iQ_5(\sigma
+\chi\varphi)}{\rm e}^{-\frac{Q_5}{2}\varphi^2}Z(r)\ ,
\end{equation}
which is invariant under the $\gamma$-shift.
On the other hand, integrating \eqref{soldel}
with respect to $\delta$, one can
restore the original $U(1)$ symmetric solution \eqref{sol5sym}.
Indeed, it is easy to show that
\be
\int_0^{2\pi} \tilde\Lambda_{\delta}^{(5)}\,\frac{{\rm d}\delta}{2\pi}
= {\rm e}^{\pm iQ_5\(\sigma+\hf\,\chi\vph\)} \,{\rm e}^{-\frac{Q_5}{4}
\(\chi^2+\vph^2\)} Z(r) =\Lambda^{(5)}.
\label{solsym}
\ee
Thus, the solutions \eqref{sol51ex} and \eqref{sol5sym} generate two equivalent bases of
instanton corrections and the integral transformations allow to pass
from one to the other. Notice that the integral transform
\eqref{solsym} maps the angle variable $\sigma$ to the scalar $D$
dual to the NS 2-form (see \eqref{Dsigma}).

It is instructive to discuss the fate of the perturbative isometries
for each particular solution. Every ${\tilde
\Lambda}_{\delta}^{(5)}$ is invariant with respect to discrete
shifts of $\sigma$ and continuous shifts along a particular
direction in the $\chi$-$\vph$ plane (accompanied by a compensating
transformation of $\sigma$), whereas the rotation symmetry \eqref{HMis} is
broken completely. In particular ${\tilde \Lambda}^{(5)} = 
{\tilde \Lambda}_{\delta = 0}^{(5)}$ preserves the shift symmetry 
in $\varphi$, and therefore falls into 
class (A) as defined in \eqref{isom}. In contrast the solution $\Lambda^{(5)}$
as in \eqref{sol5sym} is symmetric in $\chi$ and $\vph$ and respects
the U(1) isometry corresponding to \eqref{HMis}, whereas all
continuous shifts are broken.
Thus all our basis five-brane solutions preserve a
residual symmetry group is $\bZ\times \bR$ where the discrete factor
corresponds to shifts of $\sigma$ and $\bR$ comes from either the
rotation or the shift symmetry in the $\chi$-$\vph$ plane. 
Linear combinations of these basis solutions will, however, generically
break this residual symmetry to $\bZ$.

The question that now remains is which solution corresponds to the
physically realized one in non-perturbative string theory. From our
analysis, all solutions are on equal footing and preserve and break
the same amount of isometries, although the anomalies appear in
different sectors. However, these solutions yield different
metrics and hence different low-energy effective actions.
A possibility is that this also happens in string
theory. From quantum field theory we know that, in the presence of
more than one (classical) symmetries, it is possible to move
anomalies from one current to another, depending on how one quantizes
the theory and which regularization scheme is chosen. It is
conceivable that such a mechanism also works in non-perturbative
string theory. In that case, all our solutions could be considered as
physically equivalent. It would be very interesting to understand this
mechanism in more detail.

\subsection{Membrane instantons}

Since in this case the instanton action scales like $\gst^{-1}$, we
have the following ansatz
\be
\Lambda^{(2)}=A_0
r^{\alpha}\exp\(-f_{1/2}\sqrt{r}+
\sum_{k\in\Nint/2}^{\infty}\frac{A_{k}}{r^{k}}\).
\label{memans}
\ee
The master equation \eqref{Leq} is solved in appendix \ref{C} where
the following two {\it exact} solutions are found
\beq
\Lambda^{(2)}_1 &=&
\frac{C}{r} \, {\rm e}^{i \Qch \chi+i \Qph \vph} \, K_0(2
Q_2\sqrt{r+c})\ ,
\label{eq:2memr}
\\
\Lambda^{(2)}_2 &=& C'\, \frac{ \(\sqrt{1+\frac{\chi^2}{
4(r+c)}}-\frac{\chi}{ 2\sqrt{r+c}}\)^{2cQ_5}}{
r\sqrt{4(r+c)+\chi^2}} \, {\rm e}^{\pm iQ_2\vph\mp
iQ_5\sigma}\nonumber\\
&&\quad\times \exp\[-\(Q_2+\frac{\chi}{2}\,Q_5\)\sqrt{4(r+c)+\chi^2} \]\ ,
\label{solmemex}
\eeq
where $K_0$ is the modified Bessel function
and we require $Q_2$ and $\chi Q_5$ to be strictly positive. In
the first solution the charges are related by \be Q_2^2=\Qch^2 +
\Qph^2\ . \label{relch} \ee The leading terms in the expansion in
powers of $\gst$ for the two instanton corrections are
\beq
\Lambda^{(2)}_1 &\approx& \frac{C}{ r} \, {\rm e}^{i \Qch \chi+i
\Qph \vph} \exp\[-2Q_2\sqrt{r}\],
\\
\Lambda^{(2)}_2 &\approx& C' r^{-3/2}\,{\rm e}^{\pm iQ_2\vph\mp
iQ_5\sigma} \exp\[-(2Q_2+\chi Q_5)\sqrt{r}\].
\label{solmemy}
\eeq
By further expanding in powers of $\gst$, one generates the loop 
expansion around the membrane instanton. Notice that the two solutions
have a different leading power of $r$ in front of the exponent. This
power is fixed by extending the leading order solution to the full 
one-instanton result, and depends crucially on the presence of $Q_5$.

The first solution \eqref{eq:2memr} reproduces the leading behavior
of the instanton corrections found in \cite{BBS} and in \cite{DSTV}.
It depends on two charges associated with both RR fields. However,
based on the analysis \cite{DSTV}, which goes beyond the linear
order in $\Lambda$, one would expect that continuation of the
solution to multi-instanton sectors will require setting one of the
charges in \eqref{eq:2memr} to zero.

The second solution \eqref{solmemex} nicely coincides with the
result \eqref{meminst} based on the supergravity analysis for the
computation of the instanton action. It also depends on two charges,
$Q_2$ and $Q_5$, but their origin is different. For $Q_5=0$, the solution
preserves the shift symmetry in $\sigma$ and therefore belongs to class
(B) in \eqref{isom}. For $Q_5 \neq 0$, we observe that the
second charge gives rise to the factor ${\rm e}^{\pm iQ_5\sigma}$,
which is the same as in the fivebrane instanton solution. Thus, it
is natural to expect that the charge $Q_5$ does have its origin in the
NS-fivebrane, as was proposed in \cite{DdTV}. This fact also
explains why the corrections including this $Q_5$ charge were not
found in \cite{DSTV}. The reason is that the application of the Toda
equation requires the presence of the isometry in $\sigma$, whereas
non-vanishing $Q_5$ necessarily breaks it.

An interesting feature is that the instanton solution
\eqref{solmemy} is defined only for the phases (the theta-angle like
terms) of a particular relative sign.
Since for $\chi>0$ both $Q_5$ and $Q_2$ should be positive,
the phases should be of different signs in this case. Correspondingly,
in the opposite case of $\chi<0$, the charges must have opposite sign and
the phases are of the same sign. This indicates that only a
configuration of an instanton and (anti-)instanton is stable for
$\chi<0$ ($\chi>0$) and thus the stability depends on the sign of
the RR field.

\subsection{Instanton corrected metric}

Once $\Lambda$ is found, one can determine the corrections to the
hypermultiplet metric \eqref{HMq}. In terms of the real coordinates
\eqref{change} the Przanowski metric takes the form
\beq
{\rm d}s^2&=&\(\p_u^2 h+ \p_{\sigma}^2 h\)\({\rm d}u^2+{\rm d}\sigma^2\)
+\(\p_{\chi}^2 h +\p_{\vph}^2 h+2{\rm e}^h\)\({\rm d}\chi^2+{\rm
d}\vph^2\)
\nonumber \\
&& +2\(\p_{\chi}\p_u h+\p_{\vph}\p_{\sigma} h\)\({\rm d}u {\rm
d}\chi+{\rm d}\sigma {\rm d}\vph\)
+2\(\p_{\chi}\p_{\sigma}h-\p_{\vph}\p_u h\)\({\rm d}\sigma {\rm
d}\chi-{\rm d}u {\rm d}\vph\)\ .\qquad
\label{Przmetr}
\eeq
Plugging $u(r,\chi)$ from \eqref{uy} and the instanton corrected function $h$
from \eqref{hcor}, and keeping only the linear terms in $\Lambda$
one obtains the following result
\be
{\rm d}s^2={\rm d}s_{\rm UHM}^2+{\rm d}s_{\rm cor}^2,
\ee
where the instanton correction to
the hypermultiplet metric reads
\beq
{\rm d}s_{\rm cor}^2&=&
\(\(f^{-1}\, \p_r\)^2 \Lambda+ \p_{\sigma}^2 \Lambda\) \( f^2 {\rm
d}r^2+{\rm d}\sigma^2\)
\nonumber \\
&& +\(\p_{\chi}^2\Lambda+\(\p_\vph-\chi\p_{\sigma}\)^2\Lambda
+f^{-1}\,\p_r\Lambda+\frac{2(r+c)}{r^2}\,\Lambda \){\rm d}\chi^2
\nonumber \\
&& +\(\(\p_{\chi}+\chi\,f^{-1}\,\p_r\)^2\Lambda+\p_\vph^2\Lambda
+\frac{2(r+c)}{r^2}\,\Lambda \) {\rm d}\vph^2
\nonumber \\
&& +2\(\p_{\chi}\p_r \Lambda+ f\,\(\p_{\vph}-\chi \p_{\sigma}\)
\p_{\sigma} \Lambda\){\rm d}r {\rm d}\chi
\vphantom{\Bigr)}
\nonumber \\
&& +2\(\(\p_{\chi}+\chi\,f^{-1}\,\p_r\)f^{-1}\,\p_r \Lambda
+\p_{\vph}\p_{\sigma} \Lambda\){\rm d}\sigma {\rm d}\vph
\vphantom{\Bigr)}
\nonumber \\
&&+2\(\p_{\chi}\p_{\sigma}\Lambda-f^{-1}\(\p_{\vph}-\chi\p_{\sigma}\)\p_r
\Lambda\) \(( {\rm d}\sigma+\chi {\rm d}\vph ){\rm d}\chi -f \,{\rm
d}r {\rm d}\vph\), \vphantom{\Bigr)}
\label{dscor}
\eeq
and
\be
f \equiv \frac{r+2c}{r+c}\ .
\ee
The complete hypermultiplet
moduli space metric including membrane and fivebrane instantons is
based on taking the sum $\Lambda = \Lambda^{(5)}+\Lambda^{(2)}$, at
least in the one-instanton approximation.

We can also consider the cases of membranes and fivebrane instantons
separately. In \cite{Davidse:2004gg} fivebrane instanton
corrections were explicitly computed using the four-dimensional
effective supergravity action as a microscopic theory. Such an
approach has of course its limitations, since one cannot compute the
fluctuations around the instantons within supergravity. However, the
results for the instanton action and correlation functions can be
expected to give a reliable answer in the semi-classical
approximation, with the one-loop determinants left unspecified. The
metric obtained in this way was presented as
\be
{\rm d}s^2=\frac{1}{r^2} \,{\rm d}r^2+
\frac{1}{r}\((1-Y) {\rm d}\chi^2
-2i \tilde Y {\rm d}\chi {\rm d}\vph +(1+Y){\rm d}\vph^2\)+
\frac{1}{r^2} \( {\rm d}\sigma+\chi {\rm d}\vph\)^2,
\label{HMY}
\ee
where in our notations
\be
Y=Y_+ +Y_-\ , \qquad \tilde Y=Y_+ -Y_- \
,\qquad Y_{\pm}=\frac{1}{16\pi^2}\, {\rm e}^{\pm i Q_5\sigma}\,
\(S_{\rm inst}^{(5)}\)^2 K_{\rm 1-loop}\,{\rm e}^{-S_{{\rm inst}}^{(5)}}\ .
\label{YYY}
\ee
Here, $S_{{\rm inst}}^{(5)}$ is
the real part of the fivebrane instanton action \eqref{fiveinst} 
and $K_{\rm 1-loop}$ is the one-loop determinant which remained unknown. 

To show that this result agrees with the one obtained in \eqref{dscor}, let
us choose in \eqref{dscor} $\Lambda=\Lambda^{(5)}_+ +\Lambda^{(5)}_-$ where
$\Lambda^{(5)}_{\pm}$ denote the fivebrane solutions \eqref{sol51ex}
with plus and minus signs in the $\sigma$-dependent exponent,
respectively. Then the instanton correction to the metric in the
leading approximation is
\beq
{\rm d}s_{\rm cor}^2&\approx&2 Q_5\(
\Lambda^{(5)}_+ +\Lambda^{(5)}_-\) \( -{\rm d}\chi^2- (1-2 Q_5
\chi^2){\rm d}\vph^2 + 2Q_5 \chi
\({\rm d}\sigma {\rm d}\vph+{\rm d}r {\rm d}\chi\)\)
\nonumber \\
&&  - 2i Q_5  \chi \( \Lambda^{(5)}_+ -\Lambda^{(5)}_-\)
\({\rm d}\sigma {\rm d}\chi-{\rm d}r
{\rm d}\vph+\chi {\rm d}\chi {\rm d}\vph\)\ .
\label{dscor55}
\eeq
It is easy to show that the
following change of variables
\be
\chi\to \chi + 2Q_5 r \chi  \(
\Lambda^{(5)}_+ +\Lambda^{(5)}_-\), \qquad \vph\to \vph + 2iQ_5 r
\chi  \( \Lambda^{(5)}_+ -\Lambda^{(5)}_-\) \label{changev}\ ,
\ee
brings the metric \eqref{dscor55} to the form
\be
{\rm d}s_{\rm cor}^2
\approx 2 Q_5\(2Q_5\chi^2-1\) \[ \( \Lambda^{(5)}_+
+\Lambda^{(5)}_-\)\({\rm d}\vph^2  -{\rm d}\chi^2\) -2i \(
\Lambda^{(5)}_+ -\Lambda^{(5)}_-\) {\rm d}\chi {\rm d}\vph \]\ .
\ee
Upon identification
\be
Y_{\pm}=2
Q_5\(2Q_5\chi^2-1\)r\,\Lambda^{(5)}_{\pm},
\label{identY}
\ee
one finds precise agreement with the instanton corrections to the
metric in \eqref{HMY}. Due to the leading behavior
\eqref{leadsol5}, the identification \eqref{identY} is compatible
with \eqref{YYY}. Moreover, it allows us to read off the
behavior of the one-loop determinant, up to a numerical constant,
\be
K_{\rm 1-loop}= 64\pi^2 C\,
\gst^{6+2cQ_5}\(\chi^2-\frac{1}{2Q_5}\) .
\ee

One can do a similar analysis for membrane instantons, by plugging
in the solution \eqref{eq:2memr} or \eqref{solmemex} into the metric
\eqref{dscor}. Since in this case, we have no supergravity (or
string theory) calculations to compare with, we refrain from giving
explicit formulae.

\section{Discussion}

In this paper, we have used the constraints from quaternion-K\"ahler
geometry to determine the structure of the membrane and NS-fivebrane instanton
corrections to the universal hypermultiplet moduli space. As we have
shown, the constraints reduce to solving a non-linear differential
equation in terms of a single function. To describe the one-instanton
corrections, including the perturbative fluctuations around it, it is 
sufficient to find solutions of the linearized differential equation, which 
is what we did in this paper. The solutions that we presented
still contain undetermined integration constants.
To fix these constants, one presumably needs to do a microscopic string theory
calculation like the computation of the one-loop determinant around the 
membrane or fivebrane instanton.

To go beyond the one-instanton sector, one needs to solve the full non-linear
differential equation. This can be done by setting up an iteration scheme
similar to the one described in \cite{DSTV}.
Most ideally the entire instanton series sums up to some special
function respecting the symmetries of non-perturbative string
theory compactified on Calabi-Yau threefolds. 
This is similar in spirit as to how modular functions in ten-dimensional IIB
supergravity effective actions arise and determine the contributions
from D-instantons by imposing $SL(2,\bZ)$ symmetry non-perturbatively
\cite{Green:1997tv}. We leave this for future investigation.

\section*{Acknowledgements}

It is a pleasure to thank M. Ro\v{c}ek and U. Theis for stimulating
discussions. The work of S.A. is supported by CNRS. F.S. is supported by
a European Commission Marie Curie Postdoctoral 
Fellowship under contract number MEIF-CT-2005-023966. This work is partly
supported by NWO grant 047017015, EU contracts MRTN-CT-2004-005104 and 
MRTN-CT-2004-512194, and INTAS contract 03-51-6346.

\appendix

\section{Generic form of an instanton correction}
\label{D}

This appendix discusses some generic properties of instanton
corrections to the universal hypermultiplet which arise as a
consequence of the master equation \eqref{Leq}. In particular it is
established that the master equation does not allow for solutions
where the leading $g_s$-dependence of the instanton action is of the
form $S_{\rm inst} \propto 1/g_s^n, n \ge 3$. Furthermore it is
shown that there are no solutions satisfying the ``instanton
conditions'' (1) - (5) which simultaneously include non-vanishing
$1/g_s^2$ and $1/g_s$-terms in the instanton action. The later
property implies that the three-charge instanton actions discussed
in \cite{GS1} do not give rise to supersymmetric corrections to the
universal hypermultiplet moduli space.

Our starting point is the ansatz \eqref{genans}
\be
\Lambda= A_0 \, r^{\alpha} \exp \left(-\sum_{k=1/2}^p f_{k} r^{k}
+\sum_{k\in\Nint/2}^{\infty}\frac{A_{k}}{r^{k}} \right) \, ,
\ee
where, according to our general discussion, $p = \Nint/2$ is finite
and all $f_k$, $A_k$ except $A_0$ are real and $\sigma$-independent.
We then substitute this ansatz into the master equation \eqref{Leq}
and expand the resulting l.h.s.\ divided by $\Lambda$ in inverse
powers of $r$. This gives a system of differential equations on the
coefficients $f_k$ and $A_k$. Let us solve them one by one taking
into account the conditions listed in section \ref{masterinst}.

\subsection{Leading $g_s$ dependence of the instanton action}

We start by proving that there are no solutions where the leading
$g_s$-dependence of the instanton action is of the form $S_{\rm
inst} \propto 1/g_s^n, n \ge 3$.

Let us assume that $p>1$. Then the first non-trivial equation
appears at the order $r^{2p}$ and reads
\be\label{eq:A.2}
(\p_{\chi} f_p )^2+(\p_{\vph} f_p )^2=0.
\ee
This implies that $f_p$ is a constant.
Taking this into account, one finds the next equation at the order $r^{2p-1}$
\be
p^2 f_p^2+
(\p_{\chi} f_{p-\hf} )^2+(\p_{\vph} f_{p-\hf} )^2=0.
\label{eqvan}
\ee
Since all $f_k$ are real, the only solution of this equation is $f_{p-\hf}=const$
and $f_p=0$. By induction this implies that all $f_k$ with $k>1$ must vanish.

The case $k=1$ is special. In this case eq.\ \eqref{eq:A.2} still
holds, implying that $f_1$ is constant. The analog of \eqref{eqvan},
however, is modified according to \be A_0^{-1} \p_{\sigma}^2
A_0+f_1^2 +(\p_{\chi} f_{1/2} )^2+(\p_{\vph} f_{1/2} )^2=0 \, ,
\label{eqf} \ee and has solutions for non-vanishing $f_1$. Thus the
master equation restricts the ansatz for the instanton corrections
to the form
\be
\Lambda= A_0 \, r^{\alpha} \, \exp\(- f_1 r-f_{1/2} \sqrt{r} +\sum_{k\in
\Nint/2}^{\infty}{A_{k}\over r^{k}} \) \, , 
\label{genansatz}
\ee
with $f_1$ being a positive constant.

\subsection{Absence of combined membrane and fivebrane instantons}

Now we want to prove that there are no solutions satisfying our
conditions (1) to (5) where $f_1$ and $f_{1/2}$ are both
non-vanishing. As we already know, $f_1$ must be a (positive)
constant, say $f_1=|Q_5|$. Then the general solution of \eqref{eqf}
can be found by separation of variables and is based on
\be\label{eq:solA0} A_0(\chi,\vph,\sigma)=\tilde A_0(\chi,\vph) \,
{\rm e}^{i\Qs\sigma} + \hat A_{0}(\chi,\vph) \, {\rm
e}^{-i\Qs\sigma} \, , \ee
where\footnote{In the following we restrict ourselves to $\hat
A_0(\chi,\vph) = 0 $, which corresponds to considering perturbations
due to instantons only. Anti-instantons can be added at any stage by
requiring the reality of the solution.}
\be
\Qs = \sqrt{(Q_5)^2 + (\p_\chi f_{1/2})^2 + (\p_\vph f_{1/2})^2 } \, .
\ee

In the next step one establishes that the $\sigma$-independence in
the coefficients $A_k$, $k>0$ requires that $\Qs$ is constant. To
this end one observes that when substituting the solution
\eqref{eq:solA0} into the equations arising at subleading powers in
$r$ these equations become polynomials in $\sigma$ with
$\sigma$-independent coefficients. In order for the solution to be
consistent, all these coefficients have to vanish separately. In
particular considering the coefficient multiplying the
$\sigma^2$-term at order $r^{0}$ one obtains:
\be
\left( \p_\chi \left( (\p_\chi f_{1/2})^2 + (\p_\vph f_{1/2})^2 \right) \right)^2
+ \left( \p_\vph \left( (\p_\chi f_{1/2})^2
+ (\p_\vph f_{1/2})^2 \right) \right)^2 = 0 \, .
\ee
As a result, $f_{1/2}$ must satisfy
\be
(\p_{\chi} f_{1/2} )^2+(\p_{\vph} f_{1/2} )^2=\Qf^2 \, ,
\label{eqfQ}
\ee
where the constant $\Qf^2$ is fixed by \eqref{eqf} as
\be
\Qf^2 =  \Qs^2-Q_5^2 \, .
\ee

Eq.\ \eqref{eqfQ} can be solved using the following trick. Introduce
a function $y(\chi,\vph)$ such that
\be
\p_{\chi}f_{1/2}=\Qf\cos y,
\qquad \p_{\vph}f_{1/2}=\Qf\sin y.
\label{derf}
\ee
This solves \eqref{eqfQ} but the function $y$ must satisfy the integrability
condition \be \cos y\,\p_{\chi}y +\sin y\,\p_{\vph}y=0.
\label{chivphy} \ee The general solution of this equation fits into
two classes where $y$ is either a non-trivial function of $\chi$ and
$\vph$ or simply a constant. We consider these two cases in turn.

\subsubsection*{The case $y \ne const$}

In this case solution of \eqref{chivphy} can be written in an
implicit form \be \varphi\cos y-\chi\sin y=F(y), \label{difsol} \ee
where $F(y)$ is an arbitrary function. The corresponding function
$f_{1/2}$ is \be f_{1/2}=2Q_2+\Qf\(\chi \cos y+\varphi\sin y -\int
F(y)\,dy\). \label{fhalfd} \ee

Let us show that this solution
is not consistent with our conditions on the instanton corrections.
For this we go to the next constraint, which arises at the order $r^{1/2}$:
\be
 (|Q_5| - \p_\chi^2 - \p_\vph^2) f_{1/2} -
2\(\p_{\chi}f_{1/2}\,\p_{\chi}+\p_{\vph}f_{1/2}\,\p_{\vph}\)\log A_0
+2i\Qs\chi\p_{\vph}f_{1/2}=0.
\label{rhalf}
\ee
First note that for $\Qf=0$, $f_{1/2} = 2 Q_2$ is just a constant
and \eqref{rhalf} requires $Q_2 = f_{1/2} = 0$. This is, however,
not consistent with our requirement that both $f_{1}$ and $f_{1/2}$
are non-zero. Thus we take $\Qf \not = 0$ in the following. In this
case eq.\ \eqref{rhalf}  can be used to determine $A_0$.

It is convenient to use the independent variables $y$ and $\chi$ instead of $\vph$ and $\chi$.
Using eqs.\ \eqref{derf}, \eqref{fhalfd} and \eqref{difsol} it then follows that
\be
\begin{split}
2\Qf\cos y\, \p_{\chi}\log A_0= & \, 2|Q_5|Q_2+
\Qf |Q_5|\({\chi \over \cos y}+F(y)\tan y -\int F(y)\,dy\) \\
& \, +2i \Qf \Qs \chi \sin y - \frac{{\tilde Q}_5 \cos y}{\chi +F(y) \sin y
+ F'(y) \cos y}\ ,
\end{split}
\ee
where the derivative with respect to $\chi$ is taken at constant $y$.
>From this result one immediately concludes that the three charges $Q_5$, $\Qf$
and $Q_2$ cannot all be independent. To avoid singularities
at vanishing $\Qf$, $Q_2Q_5$ should either vanish or be proportional to $\Qf$.
Assuming that one of this situations is realized, one finds $A_0$
\beq
A_0=
{A(y) \, {\rm e}^{i\Qs\(\sigma+\hf\, \chi^2\tan y\)}
\over \big(\chi +F \sin y + F' \cos y\big)^{1/2}}\,
\exp\[{\tQ_2\chi\over \cos y}+
{|Q_5|\chi\over 2\cos y}\({\chi \over 2\cos y}+ \tilde F(y) \)\],
\label{A0y}
\eeq
where $A(y)$ is undetermined function and we introduced $\tQ_2={Q_2|Q_5|\over \Qf}$ and
\be
\tilde F(y)=F(y)\tan y-\int F(y)\, dy.
\ee

To make further conclusions, one has to consider the equation at the order $r^0$
\beq
& -\Qs^2\(\chi^2+3c\)+{1\over 4}\, f_{1/2}^2+c Q_5^2-2\alpha |Q_5|
-2\(\p_{\chi}f_{1/2}\,\p_{\chi}+\p_{\vph}f_{1/2}\,\p_{\vph}\)A_{1/2}
& \nonumber \\
& + A_0^{-1}\(\p_{\chi}^2+\p_{\vph}^2\)A_0-2i\Qs\chi\p_{\vph}\log
A_0 -3|Q_5|=0. & \label{rnul}
\eeq
{}From this equation one can
determine the coefficient $A_{1/2}$. However, since $A_{1/2}$ is
real, the imaginary part of the equation imposes conditions on the
functions introduced above. One can show that for non-vanishing
$\Qs$ ($\Qs=0$ leads to $\Qf=Q_5=0$) it vanishes
only if\footnote{For a special solution
\be
F(y)=\vph_0 \cos y-\chi_0\sin y,
\ee
where $\chi_0$, $\vph_0$ are some constants,
there is additional possibility to add in the exponent of \eqref{Ayyy}
the term $iQ_y y$ with $Q_y$ being a constant. However, such a term
does not affect any further conclusions.
}
\be
A(y)=C{\sqrt {\cos y\vphantom{A}}}\,\exp\({|Q_5|\over 4}\,\tilde F^2(y)
+\tQ_2 \tilde F(y)-{i\Qs \over 2}\(F^2\tan y+\int \((F')^2-F^2\)dy\)\).
\label{Ayyy}
\ee

Plugging this result into \eqref{A0y}, one finds
\be\label{A.22}
A_0=
{C \,{\rm e}^{i\Qs\(\sigma+\hf\,\(\tan y (\chi^2-F^2)-\int ((F')^2-F^2)dy\)\)}
\over\big({\chi\over \cos y} +F \tan y + F'\big)^{1/2}}\,
{\rm e}^{\tQ_2\({\chi\over \cos y}+\tilde F\)
+{|Q_5|\over 4}\,\({\chi\over \cos y}+\tilde F\)^2}.
\ee
One observes that due the sign of the last term in the exponent of $A_0$
the instanton correction diverges in the region where
${\chi\over \cos y}+\tilde F$ is large. Such a region can be achieved,
for example, by keeping $y$ fixed and considering large $\chi$.
Therefore, we are in contradiction with condition 1 from our list.
This implies that $Q_5$ must vanish
in agreement with our statement that either $f_1$ or $f_{1/2}$ vanish.

\subsubsection*{The case $y = const$}

There still remains a possibility to have another solution of
\eqref{chivphy} which is $y=const$. It gives
\be
f_{1/2}=2Q_2+\Qf(\chi \cos y+\varphi\sin y).
\label{fhal}
\ee
However, one can use the rotation isometry \eqref{HMis} and the
comment in the end of section \ref{masterinst} to put $y=0$. Thus,
in this case it is enough to consider
\be
f_{1/2}=2Q_2+\Qf\chi.
\label{fhalf}
\ee
Notice that it fits to the more general solution
\eqref{fhalfd} where however $y$ should be considered as a constant
instead of to be determined by \eqref{difsol}.

To investigate this type of solution further, one can proceed as in
the previous case. Then eq. \eqref{rhalf} and regularity assumptions again
imply that either
$Q_2Q_5$ vanish or $\sim \Qf$ and
\be\label{A.25}
A_0=C\, {\rm e}^{i\Qph\vph} \, {\rm e}^{i\Qs\sigma+\tQ_2\chi+{1\over 4}\, |Q_5|\chi^2}.
\ee
The $\vph$-dependence thereby follows from the vanishing of the imaginary part of \eqref{rnul}, yielding $A(\vph)=C\, {\rm e}^{i\Qph\vph}$. Thus, again for non-vanishing
$Q_5$ the solution diverges for large $\chi$ and should be
discarded. As a result, only instanton corrections
where either $f_1$ or $f_{1/2}$ are non-vanishing are admissible.

\section{Solution of the master equation in the five-brane case}
\label{B}

We now proceed and solve eq.\ \eqref{Leq} with boundary conditions
corresponding to a five-brane instanton. In order to arrive at the exact
solution \eqref{sol5sym} we thereby follow a two step procedure. In the
first step we substitute the ansatz \eqref{fiveans} into the master equation
for instanton corrections and expand the resulting expression in  
inverse powers of $r$. Equating the coefficients in this expansion
to zero leads to partial differential equations for the functions
$A_k$ which can be solved order by order. Thereby it turns out
that the dependence of the $A_k$ on $\vph$ and $\chi$ can be deduced from the
first few orders in the perturbative expansion. This knowledge allows to refine
the ansatz \eqref{fiveans} in such a way that it results in a partial differential
equation governing the dilaton dependence of the solution. The solution of this equation
 then gives rise to the exact one-instanton correction which includes all orders of 
perturbation theory around the instanton.

We start by adapting the results of the previous section to the case of
fivebrane instantons by setting $f_{1/2} = 0$. Eq.\ \eqref{eqf} then implies that
\footnote{In principle one should also include a term proportional to $\exp(-iQ_5 \sigma)$ in $A_0$. Since the final solution for $\Lambda$ has to be real, these terms can be restored by adding the complex conjugate of the solution found from $A_0$ given below.}
\be
f_1=|Q_5|, \qquad A_0=\tilde
A_0(\chi,\vph){\rm e}^{iQ_5\sigma}.
\label{solAs}
\ee
The function $\tilde A_0$ is determined by eq.\ \eqref{rnul} with $f_{1/2} = 0$ and $Q_\sigma = Q_5$:
\be
(\p_{\chi}^2+\p_{\vph}^2)\tilde A_0 -2i Q_5\chi\p_{\vph} 
\tilde A_0-(\chi^2+2c) Q_5^2 \,  \tilde A_0 -(2\alpha +3)|Q_5| \, \tilde A_0=0.
\label{rzero}
\ee
One class of solutions to this equation can be found by substituting 
$\tilde A_0^{\pm} ={\rm e}^{\pm \hf  Q_5\chi^2}\hat A_0^{\pm}$ 
and passing to complex coordinates $z=\hf(\chi+i\vph)$. The partial differential
equations determining $\hat A_0^{\pm}(z, \bar z)$ then become
\be
\begin{split}
+: &\  \p_{z}\p_{\bz}\hat A_0^+ +2
Q_5(z+\bz)\p_{z}\hat A_0^+ - \kappa^+ \hat A_0^+ =0,
\\ 
-:&\  \p_{z}\p_{\bz}\hat A_0^- -2 Q_5(z+\bz)\p_{\bz}\hat A_0^-
- \kappa^- \hat A_0^-=0,
\end{split}
\ee
with
\be
\kappa^{\pm} = 2 c (Q_5)^2 + (2 \alpha + 3) |Q_5| \mp Q_5 \, .
\ee 
Both equations can be solved by separation of variables. The
separable solutions are labelled by a complex eigenvalue $\lambda$
and are given by
\be\label{B.5}
\begin{split}
\hat A_{0,\lambda}^+ =& C \(\lambda+2Q_5
z\)^{\frac{\kappa_+}{2 Q_5}} {\rm e}^{-Q_5 \bz^2+\lambda \bz},
\\
\hat A_{0,\lambda}^- =& C \(\lambda-2Q_5 \bz\)^{-\frac{\kappa_-}{2 Q_5}} {\rm
e}^{Q_5 z^2+\lambda z}.
\end{split}
\ee
Since the functions $\tilde A_0^{\pm}$ resulting from \eqref{B.5} have different asymptotics at
large $\chi$ and $\vph$ they provide linearly independent solutions. The 
general solution of \eqref{rzero} is then given by linear combinations of these two families
\be
A_0 = {\rm e}^{iQ_5\sigma}\int d^2\lambda \(
a_+(\lambda)\, {\rm e}^{ \hf\, Q_5\chi^2}\hat A_{0,\lambda}^+(\chi, \vph)
+ a_-(\lambda)\, {\rm e}^{ -\hf\, Q_5\chi^2}\hat
A_{0,\lambda}^-(\chi, \vph)\) \, ,
\label{intA}
\ee
with arbitrary functions $a_{\pm}(\lambda)$. For large values of $\chi$ and $\vph$ only one of the branches in \eqref{intA} is bounded. Which of the two branches is regular thereby depends on the sign of $Q_5$ and is given by $A^-_0 (A^+_0)$ for positive (negative) $Q_5$, respectively. The instanton solution consists of the bounded term only
\be\label{intB}
A_0 = {\rm e}^{i Q_5 \sigma} \int d^2 \lambda \, a(\lambda) \, 
\left( \lambda - |Q_5| ( \chi - i \epsilon \, \vph ) \right)^{-\kappa} 
{\rm e}^{\frac{i}{2} \vph ( Q_5 \chi + \epsilon \lambda )} \, 
{\rm e}^{- \frac{|Q_5|}{4} (\chi^2 + \vph^2) + \hf \chi \lambda} \, , 
\ee
where
\be\label{B.7}
\epsilon = \mbox{sign}(Q_5)  \; , \qquad \kappa =  2 + \alpha + c |Q_5|\, .
\ee

Note that the parameter $\lambda$ appearing in the solution \eqref{intB} can be generated by the solution generating technique discussed in subsection \ref{ss:3.4}. To illustrate this, we start from the solution $A_0$ with $a(\lambda) \propto \delta(\lambda)$, 
\be
A_0=C\, (\chi- i \epsilon \, \vph)^{-\kappa} \,{\rm e}^{iQ_5\(\sigma+\hf\,\chi\vph\)} \, {\rm e}^{-\frac{1}{ 4}\, |Q_5| \(\chi^2+\vph^2\)},
\label{Azero}
\ee
with $C$ being a constant. The $\lambda$-dependent solution can then be obtained by applying the  $\beta$ and $\gamma$ shifts, eq.\ \eqref{HMisom}, to \eqref{Azero} and identifying $\lambda = - |Q_5|(\gamma - i \epsilon \beta)$. Superposing the corresponding solutions with a suitable measure yields the general solution \eqref{intB}. Based on this observation, we can then simplify our further analysis by working with the particular solution \eqref{Azero} in the following.

To get some insights on the field dependence of the subleading coefficients $A_{k}(\chi, \vph)$, $k \ge 1$ we pass to the next order in the expansion and consider the coefficient at order $r^{-1}$.
Equating this coefficient to zero results in a partial differential equation for $A_1(\chi, \vph)$ 
\be
\begin{split}
 (\p_{\chi}^2+\p_{\vph}^2) A_1+2\(\p_{\chi}\log A_0\,\p_{\chi}
+\(\p_{\vph}\log A_0-iQ_5\chi\)\p_{\vph}\)A_1 & \\
 +2|Q_5| A_1 -c^2 Q_5^2 -2c(\alpha+1)|Q_5|+(\alpha+1)^2 =0. &
\label{rminus}
\end{split}
\ee
This equation is solved by substituting $A_0$ from \eqref{Azero} and splitting the resulting complex equation into its real and imaginary part. Taking into account that $A_k(\chi, \vph)$, $k \ge 1$ are real the imaginary part yields
\be\label{minusim}
\left( Q_5 - 2 \kappa \epsilon \, (\chi^2 + \vph^2 )^{-1} \right) (\vph \p_\chi - \chi \p_\vph ) A_1(\chi, \vph) =  0 \, .
\ee
This implies that $A_1(\chi, \vph) = A_1(\rho)$, with $\rho = \chi^2 + \vph^2$. Rewriting the real part of \eqref{rminus} in terms of $\rho$ gives an ordinary differential equation for $A_1(\rho)$
\be \label{minusreal}
 4 \rho \p_\rho^2 A_1 -  2 \left( \rho |Q_5| + 2 \kappa - 2 \right) \p_\rho A_1 + 2 |Q_5| A_1  - c^2 Q_5^2 - 2 c (\alpha + 1 ) |Q_5| + (\alpha+1)^2 =  0 \, , 
\ee
 which has the solution
\be\label{B.13}
\begin{split}
A_1(\rho) = & \, \left( 2 \kappa - 2 + |Q_5| \rho \right) \, 
\left[ C_1 + C_2 \int d \rho \frac{ \rho^{\kappa-1} \, {\rm e}^{\hf |Q_5| \rho}}{(2 \kappa  - 2 + |Q_5| \rho)^2}  \right] \\
& \, - \frac{1}{2|Q_5|} \left( c^2 Q_5^2 + 2 c (\alpha +1) |Q_5| - (\alpha + 1)^2  \right) \, . 
\end{split}
\ee
In the limit $Q_5\to 0$ the solution $A_1(\rho)$ is regular for $\alpha = -1$ only. 
But since $Q_5$ is the only instanton charge of the solution, and the instanton ``does not exist'' if $Q_5 = 0$ we do not insist on the regularity of $A_1(\rho)$ at this point in accord with condition (4). In fact we will argue below that the physical instanton solution should correspond to $\kappa = 0$ and, by virtue of \eqref{B.7}, to $\alpha = -2 - c|Q_5|$.    

At higher orders of the expansion the pattern encountered for $A_1(\chi, \vph)$ repeats itself. The
partial differential equations which determine the functions $A_k(\chi, \vph)$ for $k \ge 2$ again split into a real and imaginary part. The later has to vanish independently at every order in the expansion and is given by
\be
\left( Q_5 - 2 \kappa \epsilon \, (\chi^2 + \vph^2)^{-1} \right) (\vph \p_\chi - \chi \p_\vph ) A_k(\chi, \vph) =  0 \, .
\ee
Based on these equations we conclude that
\be
A_k(\chi, \vph) = A_k(\rho) \, , \qquad k \ge 1 \, .
\ee

This result motivates the following refined ansatz for the five-brane instanton corrections
\be\label{fiveansfullr}
\begin{split}
\Lambda^{(5)}= & \, (\chi-
i \e \, \vph)^{-\kappa} \,{\rm e}^{iQ_5\(\sigma+\hf\,\chi\vph\)} \,
{\rm e}^{- \, \frac{1}{4} \, |Q_5| \,\rho} \, Z(r, \rho) \\
= & \,
\rho^{-\kappa/2} \, {\rm e}^{i \e \kappa \arctan(  \, \vph/\chi)} \,{\rm e}^{iQ_5\(\sigma+\hf\,\chi\vph\)} \,
{\rm e}^{-\frac{1}{4} \, |Q_5| \,\rho} \, Z(r, \rho) \, .
\end{split}
\ee
Substituting this ansatz into the master equation \eqref{Leq} gives rise to a partial differential equation for the unknown function $Z(r, \rho)$
\be\label{Leq5brr}
\begin{split}
& \bigg[(r+c)\p_r^2+ 4 \rho \p_\rho^2 + \(3+{2c\over r}\)\p_r -2(|Q_5| \rho + 2 \kappa -2) \p_\rho \\& \qquad  -Q_5^2\(r+3c+{c^2\over r+c}\)
-|Q_5| ( 1 - 2 \kappa) +{1\over r}\bigg]Z(r, \rho) =0 \, .
\end{split}
\ee
The general solution can be found by separation of variables. We will however
focus on a particular class of solutions, namely those for which $Z(r, \rho) = Z(r)$ is independent of $\rho$. 
In terms of the perturbative expansion this corresponds to setting the integration constants $C_i$ appearing in the coefficients $A_k(\rho)$ to zero (cfg. \eqref{B.13}). In this case \eqref{Leq5brr} simplifies to an ordinary differential equation for $Z(r)$. Setting
\be
Z(r) = r^{-1} \,  (r+c)^{c |Q_5|} \, \tilde Z(\xi(r)) \, , \qquad \xi=r+c \, , 
\ee
the resulting equation for $\tilde Z(\xi)$ takes the form
\be\label{Leq5brar}
\(\p_{\xi}^2+{q\over \xi}\,\p_{\xi}-Q_5^2-|Q_5| {q-2\kappa \over \xi}\)\tilde Z(\xi)=0,
\ee
with $q=1+2c|Q_5|$. For non-zero values of $\kappa$ the solution to this equation is given by Whittaker functions (cfg., {\it e.g.}, \cite{AS}). 

The case $\kappa = 0$ is special as in this case the $\theta$-angle like term $\propto {\rm exp}(i \e \kappa \arctan(  \, \vph/\chi))$ in \eqref{fiveansfullr} is absent. While such terms do not lead to a violation of the conditions (1) - (5) stated in subsection \ref{masterinst}, it seems very unlikely that the five-brane instanton solution actually contains such angle-terms. Therefore we will focus on $\kappa=0$ in the following. In this case the general solution of \eqref{Leq5brar} is
\be\label{gensolr}
\tilde Z(\xi)=C_1 \, {\rm
e}^{|Q_5|\xi}+C_2\,(2|Q_5|)^{q-1} \,{\rm
e}^{|Q_5|\xi}\Gamma\(1-q,2|Q_5|\xi\),
\ee
where $\Gamma(p,x) = \int_x^{\infty} {\rm e}^{-t}t^{p-1}\,{\rm d}t$ 
is the incomplete gamma function. The first term in \eqref{gensolr} increases
exponentially in $\xi = r + c$ and violates condition (1) while the second
term indeed gives rise to a solution which is exponentially suppressed
for large values of the dilaton. Consequently we set $C_1 = 0$ while keeping $C_2 = C$ as 
a free parameter. The resulting five-brane instanton correction is then
given by  
\be
\Lambda^{(5)} =
\frac{C}{r(r+c)^{c|Q_5|}}\,{\rm e}^{iQ_5\(\sigma+\hf\,\chi\vph\)}
\,{\rm e}^{|Q_5|\(r-\frac{1}{4}\,\(\chi^2+\vph^2\)\)}
\int_1^{\infty}{\rm e}^{-2|Q_5|(r+c)t}\frac{{\rm d}t}{
t^{1+2c|Q_5|}}\ .
\label{sol5exr}
\ee
This result completes the derivation of the $\Lambda^{(5)}$ given in eq.\ \eqref{sol5sym}.

\section{Solution of the master equation in the membrane case}
\label{C}

After discussing the five-brane instanton corrections associated with the instanton action 
\eqref{fiveinst} we now proceed with analyzing solutions of \eqref{Leq} corresponding to
membrane instantons. In order to obtain exact (linearized) solutions we again follow the two
step procedure of the previous section, {i.e.}, we first carry out a perturbative analysis in $g_s$ to gain some insights on the field dependence of the coefficient functions $A_k(\chi, \vph)$. This information is then used to refine the ansatz for the membrane case. Substituting this improved
ansatz into \eqref{Leq} then gives rise to a (partial) differential equation from which the exact solution can be determined. This program will lead to two types of membrane instanton corrections, 
the two charge membrane instanton corrections discussed in \cite{DSTV} and the membrane instanton 
with five-brane charge associated with the instanton action \eqref{meminst}. 

The starting point of the analysis are again the results obtained in Appendix \ref{D}. These are adapted to the membrane case by imposing
\be\label{C.1}
f_1 = |Q_5| = 0 \, , \qquad A_0 = \tilde A_0(\chi, \vph) {\rm e}^{i Q_\sigma \sigma}
\ee
with
\be\label{C.2}
Q_\sigma^2 = (\p_\chi f_{1/2})^2 + ( \p_\vph f_{1/2})^2 \, . 
\ee
Eqs.\ \eqref{C.1} and \eqref{C.2} give rise to two distinct classes of solutions corresponding to 
$Q_\sigma = 0$ and $f_{1/2}$ being a positive constant and $Q_\sigma \not = 0$ which leads to a field dependent leading coefficient $f_{1/2}$. We will now discuss these possibilities in turn.

\subsection{Two-charge membrane instantons}

We first consider the case $\Qs = 0$. Eq.\ \eqref{C.2} then implies that
\be\label{C.3}
f_{1/2} = 2 |Q_2|
\ee
is a positive constant. Substituting this result into eq.\ \eqref{rhalf},
 one finds that the coefficient equation at order $r^{1/2}$ does not
impose further restrictions. The equation determining $\tilde A_0(\chi, \vph)$ arises
at order $r^0$. Substituting \eqref{C.3} into \eqref{rnul} it becomes
\be \label{rzeromem}
(\p_{\chi}^2+\p_{\vph}^2) \, \tilde A_0 + Q_2^2 \, \tilde A_0=0 \, .
\ee
At this stage it suffices to consider special solutions of the form
\be\label{Azeromem}
\tilde A_0=  \( a_0 + \tilde{a}_0 \, \eta \) \,{\rm e}^{i\Qch\chi+i\Qph\vph}, \qquad \Qch^2+\Qph^2=Q_2^2 \, ,
\ee
with
\be \label{etadef}
\eta \equiv Q_\vph \chi - Q_\chi \vph \, ,
\ee
and arbitrary coefficients $a_0 , \tilde{a}_0$.
The general solution of \eqref{rzeromem} is then obtained by superimposing 
the solution \eqref{Azeromem} for different values of the charges $\Qch, \Qph$
associated with the two RR fields.

In order to get some information on the functions $A_{k/2}(\chi, \vph)$, 
$k \ge 1$ 
we consider the coefficient function at order $r^{-1/2}$.  Using the results 
\eqref{C.3} 
and \eqref{Azeromem} together with the reality of $A_{1/2}$ 
one obtains two equations corresponding to the real and imaginary part
of the coefficient multiplying $r^{-1/2}$. The imaginary part gives
rise to the equation
\be
\begin{split}
\, ( \Qch \, \p_\chi + \Qph \, \p_\vph )\, A_{1/2}  =  0 \, ,
\end{split}
\ee
which indicates that 
\be
A_{1/2}(\chi, \vph) = A_{1/2}(\eta) \,  
\ee
is a function of $\eta$ only. Using this result, the real part
equation becomes a ordinary differential equation for 
$A_{1/2}(\eta)$
\be
\( a_0 + \tilde{a}_0 \, \eta \) \, \( Q_2^2 \, \p_\eta^2 \, A_{1/2} - \tfrac{1}{2} \, |Q_2| \, (4 \alpha + 5 )  \) + 2 \, \tilde{a}_0 \, Q_2^2 \, \p_\eta A_{1/2} = 0 \, .
\ee
This equation is readily solved and one obtains
\be\label{C.7}
A_{1/2}(\eta) = \frac{1}{2 |Q_2|} \, \frac{4 \alpha + 5}{a_0 + \tilde{a}_0 \, \eta} \, 
\( \tfrac{1}{6} \tilde{a}_0 \, \eta^3 + \tfrac{1}{2} a_0 \, \eta^2 + a_{1/2} \, \eta + \tilde{a}_{1/2} \) \, .
\ee
Here $a_{1/2}$ and ${\tilde a}_{1/2}$ are real integration constants.

This pattern repeats itself for $A_{k/2}(\chi, \vph)$, $k \ge 2$. The partial differential equation determining $A_{k/2}$ decomposes into real and imaginary part. The equation arising from the imaginary part shows that $A_{k/2}(\chi, \vph) = A_{k/2}(\eta)$. Thus we conclude that the functions $A_{k/2}(\chi, \vph)$, $k \ge 1$ depend on $\chi, \vph$ through the combination \eqref{etadef} only.

This result motivates refining the ansatz \eqref{genans} to the form
\be \label{lambda-mem}
\Lambda^{(2)}_1 = {\rm e}^{i \Qch \chi+ i\Qph \vph} \, Z(r, \eta) \,  , 
\ee
with $Z(r, \eta)$ being an undetermined function. Substituting this ansatz
into the master equation \eqref{Leq} yields the following partial differential
equation for $Z(r, \eta)$
\be\label{C.10}
\[ (r+c) \, \p_r^2 + Q_2^2 \, \p_\eta^2 + \( 3 + \frac{2c}{r} \) \, \p_r - Q_2^2 + \frac{1}{r} \] Z(r, \eta) = 0 \, . 
\ee
The general solution can again be found by separation of variables $Z(r,\eta)=
f(r)g(\eta)$. The equation for $g(\eta)$ is
\be
Q_2^2 \partial_\eta^2 g = -\lambda g\ ,
\ee
for arbitrary values of $\lambda$. However, for non-zero values of $\lambda$
the resulting solutions have either unphysical boundary conditions, or
can be obtained by redefining the charges in \eqref{lambda-mem}. The 
remaining case is when $\lambda=0$ which leads to linear $\eta$ dependence
in $Z(r,\eta)$ and does not affect the $r$-dependent factor $f(r)$. 

For simplicity, we now focus on the class of solutions where $Z(r, \eta) = Z(r)$ is $\eta$-independent. The $\eta$-dependence can trivially be restored.
The exact solution of \eqref{C.10} is obtained by substituting
\be\label{C.11}
Z(r) = r^{-1} \tilde Z(\xi(r)) \; , \qquad \xi = r + c \, ,
\ee
into \eqref{C.10}. This leads to a modified Bessel equation for $\tilde Z(\xi)$
\be
( \xi \, \p_\xi^2 + \p_\xi - Q_2^2 ) \, \tilde Z(\xi) = 0 \, .
\ee
Thus
\be
Z(r) = C_1 \,r^{-1} \, I_0\(2 |Q_2| \sqrt{r+c}\) +
C_2 \,r^{-1} \, K_0\(2 |Q_2| \sqrt{r+c}\) \, , 
\ee
where $I_0$ and $K_0$ are modified Bessel functions of the second kind.
The condition that the instanton corrections are exponentially
suppressed for small string coupling requires that
$C_1 = 0$. Hence we obtain the following
exact two-charge membrane solution
\be\label{eq:2mem}
\Lambda^{(2)}_1 = \frac{C}{r} \, {\rm e}^{i \Qch \chi+i \Qph \vph} \,
K_0 \( 2 \, \sqrt{ \Qch^2 + \Qph^2 } \, \sqrt{r+c} \) \, . 
\ee
This solution corresponds to the membrane instanton corrections
discussed in \cite{DSTV}. The $\eta$-dependence can be restored by replacing
$C\rightarrow C_1 + \eta C_2$.

\subsection{Membrane instantons with fivebrane charge}

Let us now proceed and discuss the case $Q_\sigma \not = 0$.
This situation corresponds to a membrane instanton with charge $Q_2$
which may also include a five-brane charge $Q_\sigma$. Starting point
of the analysis are again the results of Appendix \ref{D} adapted to the membrane
case by imposing \eqref{C.1} and \eqref{C.2}. Considering the solution 
\eqref{fhalfd} with $\tilde Q_5 = \Qs$ we again have the two distinct
cases of $y = y(\chi, \vph)$ and $y = const$, which will now be discussed in turn.

We first show that taking $y = y(\chi, \vph)$ non-constant is not compatible 
with the requirements (1) - (5). This can be seen from
starting with the solution \eqref{A.22} and taking $Q_5 = 0$
\be
A_0=
{C \,{\rm e}^{i\Qs\(\sigma+\hf\,\(\tan y (\chi^2-F^2)-\int ((F')^2-F^2)dy\)\)}
\over\big({\chi\over \cos y} +F \tan y + F'\big)^{1/2}}.
\label{Fymem}
\ee
In the limit $Q_5 = 0$ the divergent terms in \eqref{A.22} are absent. In order
to rule out such solutions one has to proceed to the next non-trivial equation which
is given by the real part of \eqref{rnul}. Using
\be
\vrh={\chi\over \cos y} +F \tan y + F', \qquad
\Delta_F=F'+\int F\,dy,
\ee
this equation becomes
\beq
2\Qs\p_{\vrh}\,A_{1/2}(\vrh,y)&=& Q_2^2+Q_2\Qs(\vrh-\Delta_F)
-\frac{\Qs^2}{4}\(12c+\Delta_F(2\vrh-\Delta_F)\)
\nonumber \\
&&
+\frac{1}{4\vrh^{4}}\(\vrh^2-2\Delta_F''\vrh+5(\Delta_F')^2\).
\label{rnulmem}
\eeq
The solution of this equation has a singularity
in the limit $\Qs \to 0$. This contradicts the regularity condition (4). 
 Thus this type of solutions is ruled out.

We then restrict ourselves to the case where $y$ is constant.
Using the U(1) isometry
\eqref{HMis}, one can perform a rotation in the $\chi$-$\phi$-plane
such that $f_{1/2}$ depends on $\chi$ only\footnote{We assume that
we work in the region $\chi>0$.}
\be
f_{1/2}=2|Q_2|+|\Qs|\chi \, ,
\ee
Adapting the result \eqref{A.25} then leads to
\be\label{solgB}
A_0= C e^{i\Qs\sigma+i\Qph \vph} \, .
\ee

Subsequently $A_{1/2}$ is determined by the real part of \eqref{rnul}
\be\label{eqB1}
2|\Qs|\p_{\chi} A_{1/2} =\(2\Qs\Qph+|Q_2||\Qs|\)\chi
-\Qs^2\({3\over 4}\chi^2+3c\) +Q_2^2-\Qph^2 \, .
\ee
For $Q_2$ and $Q_\vph$ independent, the last two terms in this equation 
will give rise to a singularity in $A_{1/2}$ as $\Qs\to 0$ which would be in disagreement
with our condition (4). The two possibilities to avoid this singularity are either
taking $Q_2 \propto \Qs$ and $Q_\vph \propto \Qs$ or identifying
\be\label{C.12}
Q_2^2 =\Qph^2 \, .
\ee
Requiring that $Q_2$ is an independent charge which does not vanish in the
limit when $\Qs\to 0$ indicates that one should choose the condition \eqref{C.12}.
It is then straightforward to determine $A_{1/2}$ by integrating \eqref{eqB1},
\be
A_{1/2} =-{1\over 8}|\Qs|\chi^3+ {2\e+1\over 4}|Q_2|\chi^2
-{3c\over 2}|\Qs| \chi + a_{1/2}(\vph),
\label{solB1}
\ee
where $\e=\sign(Q_2\Qs)$ is the relative sign of the two charges and
$a_{1/2}$ is arbitrary function.

Subsequently, one moves to higher orders in the perturbative expansion.
Imposing the reality of the $A_k(\chi, \vph), k > 0$, each order
of the expansion gives rise to two independent equations originating
from the real and imaginary part of the coefficients. 
Using the equations coming from the imaginary part one can proof by 
induction that $A_k(\chi, \vph) = A_k(\chi), k > 0$ are independent of
$\vph$ while the $\chi$-dependence of the $A_k(\chi)$ is determined by 
the real parts of the expansion. 
In particular this result establishes that $a_{1/2}(\vph)$ appearing in
\eqref{solB1} is constant.

The function  $A_1(\chi)$ is fixed by the differential equation at order $r^{-1/2}$.
Upon substituting the previous results this equation becomes
\be\label{C.22}
2|\Qs|\p_{\chi} A_1=
-(\alpha+2)|\Qs|\chi-(2\alpha+2-\e)|Q_2|.
\ee
Similarly to the situation at the previous
order the last term induces a singularity in $\Qs$
and should vanish separately. This
fixes the parameter $\alpha$ to
\be
\alpha={\e\over 2}-1 \, .
\ee
Integrating \eqref{C.22} then yields
\be
A_1=-{2+\e\over 8}\chi^2 + a_1 \, .
\ee

At this point it is illustrative to consider one additional order in the perturbative expansion, 
as this equation will give another restriction on the charges. Making use of the previous results
the equation at order $r^{-3/2}$ determines $A_{3/2}(\chi)$ and reads
\beq\label{eqB3}
\nonumber
2|\Qs|\p_{\chi} A_{3/2} &=&
\tfrac{5}{64}\,\Qs^2\chi^4
-\tfrac{1}{8} \, |\Qs||Q_2| \,  \( 4 \e + 3 \) \, \chi^3
+ \tfrac{3}{2} \, Q_2^2 \, \( 1 + \e \) \, \chi^2
+ \tfrac{5}{8} \Qs^2 \, c \, \chi^2 \\ 
&& +|\Qs| \, \(\tfrac{1}{2} a_{1/2} - c |Q_2| \, (2 + 3 \e ) \) \, \chi 
+|Q_2| \( a_{1/2} + c |Q_2| \) \\ \nonumber
&& -\tfrac{1}{4} \( 1 + \e \) 
+ \tfrac{5}{4} \, c^2 \, \Qs^2 \, .
\eeq
Requiring the absence of singularities in the limit $\Qs \to 0$ then requires
\be
\e = -1 \; , \qquad a_{1/2}=-c|Q_2| \, .
\label{b1}
\ee
Remarkably this implies that the solutions where $Q_2$ and $\Qs$ have the same sign do not
satisfy condition (4) and only solutions with $\sign(Q_2\Qs)=-1$ are physical. 

Summarizing the results of the perturbative analysis we obtain
\beq
\Lambda^{(2)}_2 & = & C \, r^{-3/2} \,e^{iQ_2\vph+i\Qs\sigma}
\exp\biggl[-(2|Q_2|+\chi|\Qs|)\sqrt{r}
\biggr.\nonumber \\
&& \left.
- \(|Q_2| \( \tfrac{1}{4} \chi^2+c \)+ \tfrac{1}{2} \chi \, |\Qs| \, \(\tfrac{1}{4}\chi^2
+3c\)\)\sqrt{r}^{-1} + \ldots \],
\label{solmem}
\eeq
where $\sign(Q_2 \Qs)=-1$.

We now use the input from the perturbative analysis to find an exact solution for the
membrane-fivebrane case. The $\vph$-independence of the coefficients $A_k, k > 0$ thereby
motivates considering the refined ansatz
\be\label{ansexact}
\Lambda^{(2)}_2 = {\rm e}^{iQ_2\vph+i\Qs\sigma} Z(r,\chi).
\ee
Substituting this ansatz into the master equation \eqref{Leq}
 leads to the following partial differential equation for $Z(r, \chi)$
\be
\[(r+c)\p_r^2+\p_{\chi}^2+\(3+ \tfrac{2c}{r}\)\p_r-\Qs^2\(r+\chi^2+3c+\tfrac{c^2}{r+c}\)
-Q_2^2-2\chi |Q_2 \Qs|+\frac{1}{r} \]Z=0,
\label{redeq}
\ee
where the condition $\mbox{sign}(Q_2 \Qs)= -1$ was used. As in the case of the fivebrane 
and membrane instanton corrections \eqref{Leq5brr} and \eqref{C.10}, equation \eqref{redeq} is
separable and may be solved by separation of variables. Here we will, however, follow a different
strategy and use input from the perturbative analysis 
\eqref{solmem} and eq.\ \eqref{redeq} to further restrict 
the ansatz \eqref{ansexact}. In this course we make the 
following observations. First notice that
upon substituting $Z(r, \chi) = r^{-1} \tilde{Z}(r, \chi)$, \eqref{redeq} 
becomes a partial differential equation for $\tilde{Z}(r, \chi)$ 
which depends on the variable $\xi = r + c$ only. 
This implies that $\tilde{Z}(r, \chi)$ depends on $r$ through the combination $r+c$. 
Second one observes that the
terms proportional to $Q_2$ in \eqref{solmem} are the first two terms
in the expansion of the instanton action \eqref{meminst} for large values
of $r$, up to $c$-dependent terms which are 
generated by replacing $r \rightarrow r+c$ in \eqref{meminst}. Third the
perturbative result \eqref{solmem} shows
that the perturbative expansion around the instanton should start with $r^{-3/2}$.
Investigating solutions of \eqref{redeq} which obey these restrictions one
then arrives at the conclusion that $Z(r, \chi)$ should be of the form
\be
Z(r,\chi)=\frac{C}{ r \sqrt{4(r+c)+\chi^2} }\,
{\rm e}^{-|Q_2|\sqrt{4(r+c)+\chi^2}-|\Qs| f(r,\chi) },
\label{ansex}
\ee
where $f(r,\chi)$ does not depend on the charges. 

Substituting \eqref{ansex} into \eqref{redeq} leads to a differential equation for $f(r,\chi)$. This equation splits into three terms which are proportional to $\Qs^2$, $Q_2 Q_\sigma$ and $|\Qs|$, respectively. Since we assumed $f(r, \chi)$ to be charge-independent each of these terms has to
vanish separately. This gives rise to the following three equations
\beq
&& \xi \(\p_\xi f\)^2+\(\p_{\chi} f\)^2  =  \xi +\chi^2+2c+{c^2\over \xi}, \\
&& 2 \, \xi \p_\xi f+\chi\p_{\chi}f  =  \chi \, \sqrt{4 \, \xi + \chi^2}, \\[1.2ex]
&& \( 4 \, \xi + \chi^2 \) (\p_\chi^2 + \xi \, \p_\xi^2) f + \( \chi^2 \p_\xi - 2 \chi \p_\chi \) f  =  0 \, .
\eeq
Remarkably these equations are simultaneously solved by taking
\be
f=\tfrac{1}{2} \,\chi\sqrt{4 \, \xi + \chi^2}
-2c\log \[ \sqrt{\xi }^{-1} \(  \sqrt{4 \, \xi + \chi^2 }-\chi \) \] \, .
\ee
Thus the ansatz \eqref{ansex} indeed gives rise to a consistent solution of \eqref{redeq}.
The corresponding {\it exact} membrane instanton
correction is based on the instanton action \eqref{meminst} and reads
\begin{eqnarray}
\Lambda^{(2)}_2 &=& C\,\tfrac{\( \sqrt{r+c }^{\,-1} \(  \sqrt{4 \, (r+c) + \chi^2 }-\chi \) \)^{2c|\Qs|}}{ r\sqrt{4(r+c)+\chi^2}}
\, {\rm e}^{ iQ_2\vph+ i\Qs\sigma}\times\nonumber\\
&&\qquad\times\exp\[-\(|Q_2|+\tfrac{1}{2} \, \chi \, |\Qs| \)\sqrt{4(r+c)+\chi^2}
\],
\label{solmemm}
\end{eqnarray}
where $Q_2$ and $\Qs$ must have opposite signs and $\chi > 0$.

\end{document}